\documentclass[%
reprint,
amsmath,amssymb,
aps,
prx,
]{revtex4-2}

\usepackage{graphicx}% Include figure files
\usepackage{dcolumn}% Align table columns on decimal point
\usepackage{bm}% bold math
\usepackage{hyperref}% add hypertext capabilities
\usepackage{subcaption}
\usepackage[ruled,linesnumbered]{algorithm2e}
\usepackage[mathlines]{lineno}% Enable numbering of text and display math
\usepackage{amsthm}
\newtheorem{theorem}{Theorem}

\DeclareMathOperator*{\argmin}{arg\,min}

\graphicspath{{figures/}}

\begin{document}

\title{A multi-model ensemble Kalman filter \\ for data assimilation and forecasting}% Force line breaks with \\

\author{Eviatar Bach}
\email{eviatarbach@protonmail.com}
\affiliation{%
	Geosciences Department and Laboratoire de Météorologie Dynamique (CNRS and IPSL), École Normale Supérieure and PSL University, Paris, France
}%
\affiliation{Division of Geological and Planetary Sciences, California Institute of Technology, Pasadena, United States}

\author{Michael Ghil}
\affiliation{Geosciences Department and Laboratoire de Météorologie Dynamique (CNRS and IPSL), École Normale Supérieure and PSL University, Paris, France
}%
\affiliation{Department of Atmospheric and Oceanic Science, University of California at Los Angeles, Los Angeles, United States}%

\date{\today}% It is always \today, today,

\begin{abstract}
Data assimilation (DA) aims to optimally combine model forecasts and observations that are both partial and noisy. Multi-model DA generalizes the variational or Bayesian formulation of the Kalman filter, and we prove that it is also the minimum variance linear unbiased estimator. Here, we formulate and implement a multi-model ensemble Kalman filter (MM-EnKF) based on this framework. The MM-EnKF can combine multiple model ensembles for both DA and forecasting in a flow-dependent manner; it uses adaptive model error estimation to provide matrix-valued weights for the separate models and the observations. We apply this methodology to various situations using the Lorenz96 model for illustration purposes. Our numerical experiments include multiple models with parametric error, different resolved scales, and different fidelities. The MM-EnKF results in significant error reductions compared to the best model, as well as to an unweighted multi-model ensemble, with respect to both probabilistic and deterministic error metrics.
\end{abstract}

\maketitle

\section*{Plain Language Summary}
Forecasts that combine multiple imperfect models of a system are used in many fields, including the physical, natural and socio-economic sciences. In particular, data assimilation (DA), the process by which observations are integrated with model forecasts, is critical in the prediction of chaotic systems. Multi-model DA (MM-DA) unifies multi-model forecast combination and DA into a single process. Here, we significantly improve on previous formulations of MM-DA by accounting for model error, and formulate a multi-model ensemble Kalman filter appropriate for high-dimensional systems.

\section{Introduction}\label{sec:intro}

Combining multiple forecasts from imperfect models of reality can often lead to forecasts that are better than any single model. Such multi-model forecasts have been enormously successful in weather and climate prediction \cite{hagedorn_rationale_2005,krishnamurti_review_2016}; economics \cite{clemen_combining_1989}; epidemiological forecasting, including that of COVID-19 \cite{cramer_evaluation_2022}; hydrology \cite{xue_multimodel_2014,okuno_combining_2019}; tracking and navigation \cite{bar-shalom_estimation_2001}; space weather \cite{schunk_space_2016}; air quality forecasting \cite{mallet_ozone_2009,sengupta_ensembling_2020}; and numerous other application areas \cite{clemen_combining_1989,trenkler_combination_1998,fragoso_bayesian_2018}.

Data assimilation (DA) is the process of combining model forecasts with observations to obtain a state estimate of a system. DA is an essential part of forecasting in a wide variety of scientific and engineering fields, with most methods based nowadays on the Kalman filter \cite{asch_data_2016}. In the meteorological literature, the need for methods to take noisy, possibly sparse observations and produce an initial condition suitable for a numerical model has been recognized since the first numerical weather forecast \cite{panofsky_objective_1949}, and the Kalman filter was proposed for this purpose in \cite{ghil_applications_1981}. Ensemble Kalman filters (EnKFs), which approximate the evolution of the probability distribution using a Monte Carlo approach \cite{evensen_ensemble_2003}, have become popular for geophysical and other problems \cite{hamilton_ensemble_2016}.

In this paper, we consider multi-model DA (MM-DA), a generalization of DA which allows for multiple forecast models \cite{narayan_sequential_2012}. MM-DA combines multiple forecasts \emph{and} observations, bridging the literature on forecast combination with that on DA. In this paper we make several contributions to the existing literature on MM-DA: (i) we incorporate and estimate model error in MM-DA, allowing models that are less accurate to have a lower weight, and allow the weights to differ for different variables; (ii) we formulate several possible implementations of deterministic EnKFs for MM-DA (MM-EnKFs) and discuss computational issues; (iii) we provide an open-source software implementation of MM-EnKFs; (iv) we test MM-EnKF with chaotic models for DA and forecasting in various scenarios; and, finally, (v) we prove linear minimum variance optimality of MM-DA.

The paper is laid out as follows. In sections \ref{ssec:intro_comb_forecasts} and \ref{ssec:intro_mmda}, we review the literature on combining forecasts and on multi-model DA, respectively. In section \ref{sec:mmenkf}, we discuss the development and implementation of MM-EnKFs. In section \ref{sec:experiments}, we apply MM-EnKFs to chaotic systems. In section \ref{sec:conclusions}, we draw conclusions and provide an outlook, including on applications to data-driven models. Finally, in \ref{sec:blue_proof}, we prove optimality of MM-DA in the linear minimum variance sense, and in \ref{sec:Q_est} we detail the simple model error estimation method used in the numerical experiments.

\subsection{Combining multiple model forecasts}\label{ssec:intro_comb_forecasts}

\cite{bates_combination_1969} were among the first to combine multiple distinct forecasts. They weighted multiple univariate forecasts according to past performance, and showed that the combined forecast resulted in lower error. Combining forecasts has become an important topic in statistics; see the historical overview \cite{hoeting_bayesian_1999} and the bibliographies \cite{clemen_combining_1989,trenkler_combination_1998}.

The advantage of multi-model forecasts over single-model ones, at comparable total ensemble size, comes from distinct models having different model errors. The skill of the multi-model forecast will then be improved to the extent that the model errors compensate for each other \cite{hagedorn_rationale_2005}. Furthermore, when the multi-model forecast is probabilistic, these multiple model errors may lead to better spanning the true forecast uncertainty \cite{wilks_statistical_2019}. The need for weighting comes when some models have higher skill than others, implying that the former should have higher weight in the combined forecast than the latter. The general situation, however, is that one model may not be superior to all the others in all respects. More typically, some models may have superior skill in some variables, in the representation of particular processes, or at some forecast horizons. We discuss different weighting approaches in the following subsection.

\subsubsection{Weighting distinct forecasts}
Several approaches for weighting distinct model forecasts have been developed, with Bayesian model averaging \cite{hoeting_bayesian_1999} being one of the most common ones. This methodology estimates the posterior model probabilities based on past data, and assigns the models scalar weights based on these probabilities. Dynamic versions have also been developed, to allow the weights to evolve based on current conditions. Both these methods, as well as several others discussed in \cite{narayan_sequential_2012}, are limited to scalar weights.

Other methods have been developed in the context of atmospheric prediction. Here, the models are high-dimensional, and each model may produce an ensemble of forecasts that attempts to capture the predictive uncertainty. Multi-model ensembles (MMEs), where multiple models are combined into a single ensemble without weighting, are used widely in climate prediction \cite{hagedorn_rationale_2005}. Multi-model superensembles, which weight the distinct model ensembles based on weights determined by multiple linear regression, have also been widely adopted \cite{krishnamurti_review_2016}.

The Dynamic Integrated Forecast System (DICast), developed by the National Center for Atmospheric Research (NCAR), periodically nudges model weights in the direction of error decrease \cite{myers_consensus_2011}. Cross-pollination in time (CPT) uses the forecasts of each model as initial conditions for the other models, along with some pruning rule to avoid an exponential increase of trajectories with time \cite{du_multi-model_2017,schevenhoven_efficient_2017}. In a connected supermodel, each model is nudged towards the others by introducing coupling terms in the evolution equations, and the supermodel is taken as an average of these coupled models \cite{duane_introduction_2017, selten_simulating_2017}. In a weighted supermodel, the individual models are not directly connected through coupling terms; rather, the supermodel tendency is taken to be a weighted average of the individual model tendencies, and the individual models compute their tendencies based on the supermodel state \cite{schevenhoven_improving_2019}. CPT and weighted supermodels were compared by \cite{schevenhoven_improving_2019} and \cite{schevenhoven_training_2022}.

\cite{sengupta_ensembling_2020} used a Bayesian neural network to infer model weights. Sequential aggregation takes inspiration from online learning and game theory in weighting forecasts with rules that have theoretical performance guarantees \cite{mallet_ozone_2009,thorey_online_2017,gonzalez_new_2021}. Forecast weights can also be determined using Markov chain Monte Carlo \cite{dumont_le_brazidec_quantification_2021}. Several other methods were compared for meteorological applications in \cite{young_combining_2002,gerding_adaptive_2003}.

\subsubsection{DA with multiple models}

Several methods have been proposed to weight models using DA methods, in particular relying on Kalman or particle filters. \cite{anandalingam_linear_1989} first recognized that a particular Bayesian forecast combination problem was equivalent to a Kalman filter. \cite{du_multi-model_2017} used DA in addition to CPT in combining forecasts. \cite{chen_multi-model_2019} and \cite{counillon_framework_2022} used DA to synchronize distinct models by assimilating forecasts as pseudo-observations. Multiple parametric variations, or variations in physical parameterizations in an atmospheric model, have also been used in EnKFs without weighting, in order to capture the effect of model error \cite{wu_comparison_2008,houtekamer_review_2016}.

Here, we are interested in the problem of generating an optimal state estimate using multiple model forecasts \emph{and} observations. Section 10.2 in \cite{simon_optimal_2006} proposes to run a Kalman filter for each model, and estimate its conditional probability given the observations from the innovations. These probabilities are then used as weights in combining the model forecasts. This approach is similar to the interacting multiple model \cite[IMM:][]{bar-shalom_estimation_2001} filter, popular in tracking applications, and multiple model adaptive estimation \cite{akca_multiple_2019}.

\cite{xue_multimodel_2014} combines a Bayesian model averaging approach with an EnKF, by recomputing the ensemble weights as new observations arrive. \cite{coelho_ocean_2015} estimates model weights using a separate filter. \cite{otsuka_bayesian_2015} implements a multi-model EnKF by adjusting the number of ensemble members for each model at every cycle based on a Bayesian estimate of the model's probability. In the ensemble average, the model with more ensemble members is then weighted more heavily. \cite{mallet_ensemble_2010} combines the sequential aggregation approach with DA.

\subsection{Multi-model data assimilation (MM-DA)}\label{ssec:intro_mmda}
\begin{table}[b]
\begin{tabular}{p{0.22\columnwidth}p{0.75\columnwidth}}
$\mathbf{b}_m$ & bias of $m$th model \\
$\mathbf{B}$ & climatological forecast error covariance matrix \\
$\{\gamma, \delta\}$ & smoothing parameter for \{inflation, model error\} estimation \\
$\mathbf{E}^{\{\text{f},\text{a}\}}$ & \{forecast, analysis\} ensemble \\
$\mathbf{E}_{1:m}^{\text{f}'}$ & multi-model forecast ensemble after averaging over models 1 to $m$ \\
$\{\mathbf{G}_{m}, \mathcal{G}_{m}\}$ & \{linear, nonlinear\} mapping from reference model space $m_r$ to the space of model $m$ \\
$\{\mathbf{G}_{m_1\to m_2}, \mathcal{G}_{m_1\to m_2}\}$ & \{linear, nonlinear\} mapping from space of model $m_1$ to space of model $m_2$\\
$\{\mathbf{H}, \mathcal{H}\}$ & \{linear, nonlinear\} observation operator of reference model \\
$\{\mathbf{H}_m, \mathcal{H}_m\}$ & \{linear, nonlinear\} observation operator of $m$th model \\
$\mathbf{K}$ & gain matrix \\
$m_r$ & reference model \\
$M$ & number of models \\
$\{\mathbf{M}, \mathcal{M}\}$ & \{linear, nonlinear\} forecast model \\
$n_m$ & dimension of $m$th model \\
$N_m$ & ensemble size of $m$th model \\
$\bm{\rho}$ & localization matrix \\
$\mathbf{P}^{\{\text{f}, \text{a}\}}$ & \{forecast, analysis\} error covariance \\
$\mathbf{P}_{1:m}^{\text{f}'}$ & multi-model forecast error covariance after averaging over models 1 to $m$ \\
$\mathbf{Q}$ & model error covariance \\
$\mathbf{R}$ & observation error covariance\\
$\mathbf{x}^{\{\text{t}, \text{f}, \text{a}\}}$ & \{true, forecast, analysis\} state \\
$\mathbf{x}^{\text{f}'}_{1:m}$ & multi-model forecast state after averaging over models 1 to $m$ \\
$(\mathbf{x}^{\{\text{f}, \text{a}\}})_i$ & $i$th member of \{forecast, analysis\} ensemble \\
$\overline{\mathbf{x}}^{\{\text{f}, \text{a}\}}$ & mean of \{forecast, analysis\} ensemble \\
$\mathbf{X}^{\{\text{f}, \text{a}\}}$ & \{forecast, analysis\} ensemble anomalies \\
$\mathbf{y}$ & observation
\end{tabular}
\caption{Definition of symbols. The superscript convention follows \protect\cite{ide_unified_1997}.}
\label{table:glossary}
\end{table}

In this paper, we consider a generalization of the Kalman filter formulation to multiple models. This generalization differs from the methods in the previous paragraphs in the models' and the observations' weights being determined as part of the filtering process itself, instead of being estimated separately. Multi-model DA (MM-DA), proposed by \cite{logutov_multi-model_2005} and \cite{narayan_sequential_2012}, is based on the variational or Bayesian formalisms from which the Kalman filter and related methods are derived, except that multiple models are included.

The MM-DA formulation was perhaps first studied by \cite{logutov_multi-model_2005}, who also proposed an expectation maximization algorithm for estimating the forecast error parameters along with the state estimate. The connection to the Kalman filter was not explicitly made in \cite{logutov_multi-model_2005}. The same formulation was independently developed by \cite{narayan_sequential_2012}, who showed that it can be implemented by using an iterative method. We base our exposition on \cite{narayan_sequential_2012}, but using the common DA notation of \cite{ide_unified_1997}; see Table~\ref{table:glossary} for a definition of symbols.

Suppose we have $M$ models, with each model $m$ having its own forecast state $\mathbf{x}^\text{f}_m \in \mathbb{R}^{n_m}$ with forecast error covariance matrix $\mathbf{P}^\text{f}_m$. Each model is assumed to be unbiased. One has to choose a space for the multi-model forecasts to reside in; for example, a given spatial grid in the case of an atmospheric model. We take this to be the space of one of the model states---although this is not necessary---and refer to this model as the reference model $m_r$; its choice will be discussed later.

We then define the operators $\mathbf{G}_m: \mathbb{R}^{n_{m_r}} \to \mathbb{R}^{n_m}$ which map from the reference model space to the model space of model $m$ with dimension $n_m$. Clearly, $\mathbf{G}_{m_r} = \mathbf{I}$ and we assume for the moment that these operators are linear, although this assumption can be relaxed later.

We also have a $p$-dimensional observation vector $\mathbf{y}$ with observation error covariance matrix $\mathbf{R}$. We define the observation operator $\mathbf{H}: \mathbb{R}^{n_{m_r}}\to\mathbb{R}^p$, which maps from the reference model space to the observation space.

Each model's state evolution operator is denoted by $\mathbf{M}_m$, and it is also assumed to be linear for the moment. Later, the nonlinear state evolution operator will be denoted by $\mathcal{M}_m$.

\subsubsection{Variational formulation and direct solution}\label{ssec:direct}

\paragraph{The formulation.}
For a single model forecast $\mathbf{x}^\text{f}$ with covariance matrix $\mathbf{P}^\text{f}$, the variational formulation of the optimal state estimation problem defines a cost function $\mathcal{I}[\mathbf{x}]$ for a control variable $\mathbf{x}$ as
\begin{linenomath}
\begin{equation}
\mathcal{I}[\mathbf{x}] = \|\mathbf{x} - \mathbf{x}^
\text{f}\|^2_{(\mathbf{P}^\text{f})^{-1}} + \|\mathbf{H}\mathbf{x} - \mathbf{y}\|^2_{\mathbf{R}^{-1}}. \label{eq:objective_single}
\end{equation}
\end{linenomath}
Here we use the short-hand weighted-norm notation $\|\mathbf{v}\|^2_\mathbf{A}\equiv \mathbf{v}^T \mathbf{A} \mathbf{v}$ for a quadratic form with symmetric positive semidefinite matrix $\mathbf{A}$. This cost function measures the sum of the squared Mahalanobis distances of $\mathbf{x}$ from the forecast $\mathbf{x}^\text{f}$ and the observations $\mathbf{y}$. The minimizer of Eq.~\ref{eq:objective_single} is the assimilation step of the Kalman filter.

Equation~\ref{eq:objective_single} can be generalized to multiple models as
\begin{linenomath}
\begin{equation}
\mathcal{J}[\mathbf{x}] = \sum_{m=1}^M \|\mathbf{G}_m\mathbf{x} - \mathbf{x}_m^
\text{f}\|^2_{(\mathbf{P}^\text{f}_m)^{-1}} + \|\mathbf{H}\mathbf{x} - \mathbf{y}\|^2_{\mathbf{R}^{-1}}.\label{eq:objective}
\end{equation}
\end{linenomath}

Note that this generalization implicitly assumes that the forecast errors are mutually uncorrelated; see paragraph \ref{par:corr_model_err} for more details.

\paragraph{The direct solution.}

\begin{figure*}
\centering
\includegraphics{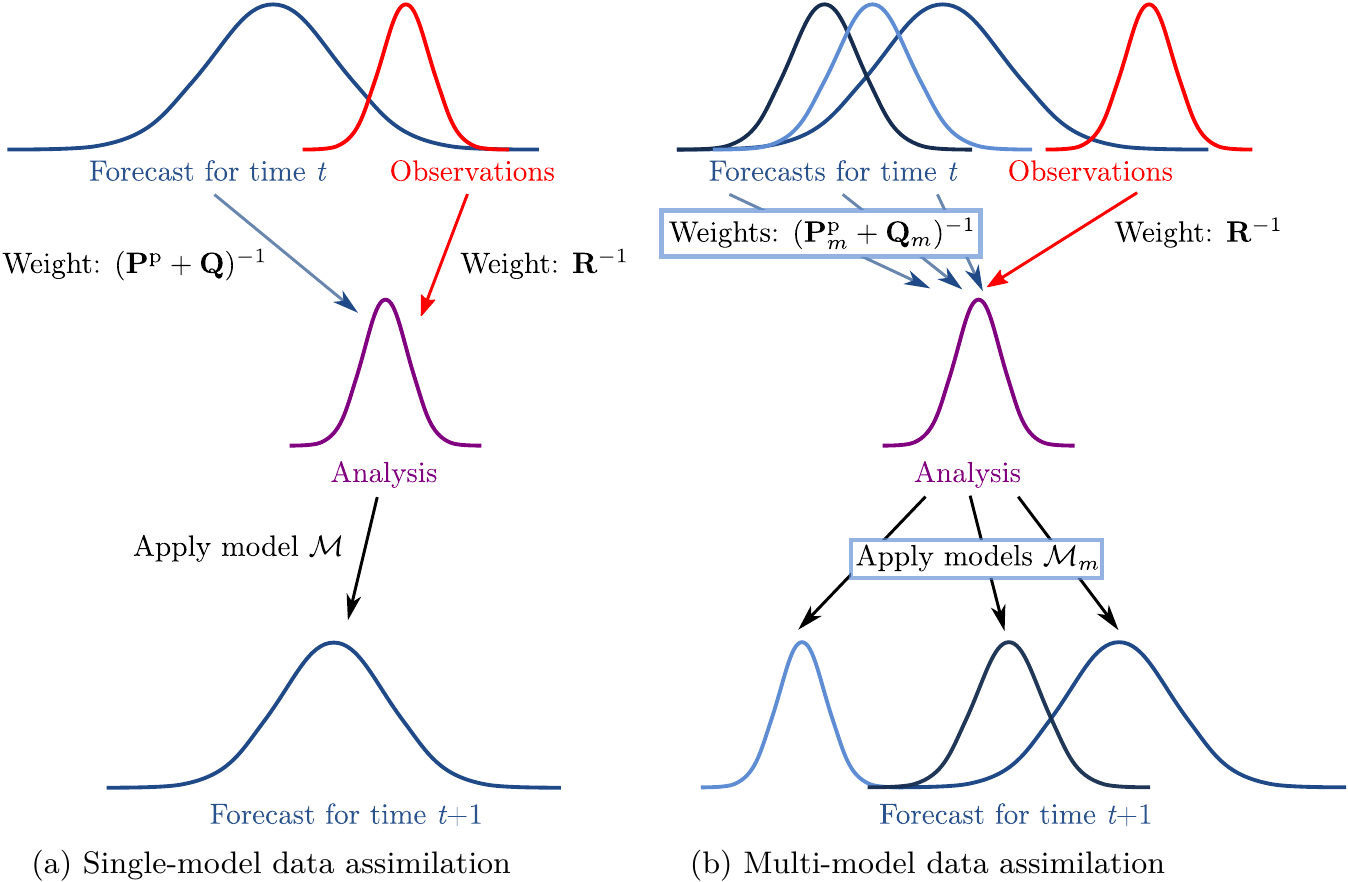}
\caption{Schematics of a forecast--assimilation cycle for single- and multi-model DA. For the purposes of this figure, we assume $\mathbf{H} = \mathbf{G}_m = \mathbf{I}$.}
\label{fig:schematic_fa}
\end{figure*}

Minimizing the multi-model cost function $\mathcal{J}[\mathbf{x}]$ above gives the analysis solution $\mathbf{x}^\text{a}$ and its corresponding covariance matrix $\mathbf{P}^\text{a}$ for the multi-model Kalman filter \cite{narayan_sequential_2012}:
\begin{linenomath}
\begin{equation}
\mathbf{x}^\text{a} = \mathbf{P}^\text{a}\left(\sum_{m=1}^M \mathbf{G}_m^T \left(\mathbf{P}^\text{f}_m\right)^{-1}\mathbf{x}^\text{f}_m + \mathbf{H}^T\mathbf{R}^{-1}\mathbf{y}\right),\label{eq:mm_xa}
\end{equation}
\end{linenomath}
where $(\cdot)^T$ is the transposition operator and
\begin{linenomath}
\begin{equation}
\mathbf{P}^\text{a} = \left(\sum_{m=1}^M \mathbf{G}_m^T \left(\mathbf{P}^\text{f}_m\right)^{-1}\mathbf{G}_m + \mathbf{H}^T\mathbf{R}^{-1}\mathbf{H}\right)^{-1}.\label{eq:mm_Pa}
\end{equation}
\end{linenomath}
The solution is thus a weighted mean, where the weights for each model $m$ are inversely proportional to $\mathbf{P}^\text{f}_m$, and the weight of the observations is inversely proportional to $\mathbf{R}$. Note that if we set $M=1$, we recover the regular Kalman filter equations.

The analysis $\mathbf{x}^\text{a}$ will be in the reference model space. The analysis in the model space for $m\neq m_r$ can then be obtained by computing
\begin{linenomath}
\begin{equation}
\mathbf{x}^\text{a}_m = \mathbf{G}_m\mathbf{x}^\text{a},\label{eq:xma}
\end{equation}
\end{linenomath}
and the analysis covariance matrix as
\begin{linenomath}
\begin{equation}
\mathbf{P}_m^\text{a} = \mathbf{G}_m \mathbf{P}^\text{a} \mathbf{G}_m^T. \label{eq:pma}
\end{equation}
\end{linenomath}

For the next forecast cycle, each model is applied to the analysis state:
\begin{linenomath}
\begin{equation}
\mathbf{x}^\text{f}_m(t_{i+1}) = \mathbf{M}_m(t_i)\mathbf{x}^\text{a}_m(t_i),
\end{equation}
\end{linenomath}
and the covariance is propagated according to
\begin{linenomath}
\begin{equation}
\mathbf{P}_m^\text{f}(t_{i+1}) = \mathbf{M}_m(t_i)\mathbf{P}^\text{a}_m(t_{i})\mathbf{M}_m(t_i)^T + \mathbf{Q}_m(t_{i}),\label{eq:cov_prop}
\end{equation}
\end{linenomath}
where $\mathbf{Q}_m$ is the model error covariance matrix for model $m$. We have introduced explicit time dependence here for clarity. In Fig.~\ref{fig:schematic_fa}, we show a schematic diagram of single- and multi-model assimilation--forecast cycles. The model error is discussed in greater detail in section \ref{sec:model_error}.

Although here we considered the variational formulation, the same equations for the multi-model Kalman filter can also be derived from the Bayesian formulation of the problem \cite{logutov_multi-model_2005,narayan_sequential_2012}.

The single-model Kalman filter is the optimal linear filter in the sense of being the minimum variance unbiased estimator. It has not previously been shown, though, that the multi-model Kalman filter is optimal in terms of minimizing variance, and we prove this in \ref{sec:blue_proof}.

\subsubsection{Iterative solution}

In some cases, it may be possible to directly compute the right-hand side of Eq.~\ref{eq:mm_xa} or to minimize Eq.~\ref{eq:objective} using an approach similar to the three-dimensional variational algorithm (3D-Var) \cite{asch_data_2016,kalnay_atmospheric_2002}. \cite{narayan_sequential_2012} show, instead, how to solve the problem iteratively.

In this iteration, the analysis of the previous model $m-1$ is considered as the forecast for the subsequent model $m$, and the forecast of model $m$ is considered as an observation:
\begin{linenomath}
\begin{subequations}\label{eq:models}
	\begin{align}
		& \mathbf{K}_m = \mathbf{P}_{1:m-1}^{\text{f}'} \mathbf{G}_m^T \left(\mathbf{G}_m\mathbf{P}_{1:m-1}^{\text{f}'}\mathbf{G}_m^T + \mathbf{P}_{m}^\text{f}\right)^\dagger, \label{eq:kalman_gain}\\
		& \mathbf{x}_{1:m}^{\text{f}'} = \mathbf{x}_{1:m-1}^{\text{f}'} + \mathbf{K}_m (\mathbf{x}_m^\text{f} - \mathbf{G}_m \mathbf{x}_{1:m-1}^{\text{f}'}), \label{eq:kalman_fcst} \\
		& \mathbf{P}_{1:m}^{\text{f}'} = (\mathbf{I} - \mathbf{K}_m\mathbf{G}_m)\mathbf{P}_{1:m-1}^{\text{f}'}; \label{eq:kalman_pa}
	\end{align}
\end{subequations}
\end{linenomath}
here $\dagger$ indicates the Moore--Penrose pseudoinverse, $\mathbf{x}_{1:m}^{\text{f}'}$ indicates the combined forecast of models 1 to $m$, and $\mathbf{P}_{1:m}^{\text{f}'}$ indicates the forecast covariance matrix of $\mathbf{x}_{1:m}^{\text{f}'}$.
Once done with the $M$ models, one assimilates the actual observations:
\begin{linenomath}
\begin{subequations}\label{eq:obs}
	\begin{align}
		& \mathbf{K} = \mathbf{P}_{1:M}^{\text{f}'} \mathbf{H}\left(\mathbf{H}\mathbf{P}_{1:M}^{\text{f}'}\mathbf{H}^T + \mathbf{R}\right)^\dagger, \label{eq:obs_assim1} \\
		& \mathbf{x}^\text{a} = \mathbf{x}_{1:M}^{\text{f}'} + \mathbf{K} (\mathbf{y} - \mathbf{H} \mathbf{x}_{1:M}^{\text{f}'}), \label{eq:obs_assim2} \\
		& \mathbf{P}^\text{a} = (\mathbf{I} - \mathbf{K}\mathbf{H})\mathbf{P}_{1:M}^{\text{f}'}. \label{eq:obs_assim3}
	\end{align}
\end{subequations}
\end{linenomath}

When the covariance matrices $\mathbf{P}_m^\text{f}$ and $\mathbf{R}$ are positive definite, the iterative solution is equivalent to the direct solution. However, unlike the direct solution, the iterative solution allows the covariance matrices $\mathbf{P}_m^\text{f}$ and $\mathbf{R}$ to be singular. The iterative solution can be shown to be independent of the order in which the models and observations are assimilated, as long as there are no inconsistent zero-variance components \cite{narayan_sequential_2012}. 

Importantly, the iterative procedure suggests a way to use single-model DA methods to estimate a solution to the multi-model DA problem. Notice that Eqs.~\ref{eq:kalman_gain}--\ref{eq:kalman_pa} and Eqs.~\ref{eq:obs_assim1}--\ref{eq:obs_assim3} are the assimilation step of a single-model Kalman filter, and thus they can be replaced by any single-model DA method.

Assume now that we have a DA method that takes as input the forecast state $\mathbf{x}^\text{f}$, forecast error covariance $\mathbf{P}^\text{f}$, observation vector $\mathbf{y}$, observation error covariance $\mathbf{R}$, and observation operator $\mathcal{H}$, and returns as output the analysis state $\mathbf{x}^\text{a}$ and analysis error covariance $\mathbf{P}^\text{a}$. Denote this function by $(\mathbf{x}^\text{a}, \mathbf{P}^\text{a}) = \mathcal{F}_{\mathrm{DA}}(\mathbf{x}^\text{f}, \mathbf{P}^\text{f}, \mathbf{y}, \mathbf{R}, \mathcal{H})$. Then, for $m=2,\ldots,M$:
\begin{linenomath}
\begin{align}
(\mathbf{x}_{1:m}^{\text{f}'}, \mathbf{P}_{1:m}^{\text{f}'}) = \mathcal{F}_{\mathrm{DA}}(\mathbf{x}_{1:m-1}^{\text{f}'}, \mathbf{P}_{1:m-1}^{\text{f}'}, \mathbf{x}_m^\text{f}, \mathbf{P}_m^\text{f}, \mathcal{G}_m), \label{eq:FDA_model}
\end{align}
\end{linenomath}
and $\mathbf{x}_{1:1}^{\text{f}'} = \mathbf{x}_1^\text{f}$, $ \mathbf{P}_{1:1}^{\text{f}'} = \mathbf{P}_1^\text{f}$. Finally,
\begin{linenomath}
\begin{equation}
(\mathbf{x}^{\text{a}}, \mathbf{P}^{\text{a}}) = \mathcal{F}_{\mathrm{DA}}(\mathbf{x}_{1:M}^{\text{f}'}, \mathbf{P}_{1:M}^{\text{f}'}, \mathbf{y}, \mathbf{R}, \mathcal{H}). \label{eq:final_approx}
\end{equation}
\end{linenomath}

Note that we allow in Eqs.~\ref{eq:FDA_model} and \ref{eq:final_approx} for possibly nonlinear operators $\mathcal{G}_m$ and $\mathcal{H}$, thus relaxing the linearity assumption on the operators $\mathbf{G}_m$ and $\mathbf{H}$, since many DA methods can deal with nonlinear observation operators. However, unless these operators are linear and the $\mathcal{F}_{\mathrm{DA}}$ function is the Kalman filter assimilation step, the solution Eq.~\ref{eq:final_approx} is only an approximation to the direct solution. Furthermore, order-independence is no longer guaranteed. A related issue occurs in serial EnKFs, wherein observations are assimilated one at a time, and localization generally introduces order dependence. \cite{kotsuki_can_2017} investigated the use of different ordering rules in this setting, and a similar investigation could be carried out for MM-DA. In our results in section \ref{sec:experiments}, we briefly explore empirically the role of the order in which the iterative solution is computed.

\section{A multi-model ensemble Kalman filter (MM-EnKF)}\label{sec:mmenkf}

As discussed in the previous section, MM-DA can potentially be used with any DA method. In this section, we describe the development and implementation of a multi-model ensemble Kalman filter (MM-EnKF). One of the advantages of EnKFs in general is that they dynamically estimate the forecast error covariance matrices, and are thus able to adapt to current conditions, or ``errors of the day'' \cite{kalnay_atmospheric_2002}. In the MM-EnKF, this flow dependence is then reflected in the weights assigned to each model and the observations in the state estimate.

For each $m$, we take its ensemble to have $N_m$ members and denote the forecast and analysis ensembles as $\mathbf{E}_m^\text{f} = [(\mathbf{x}_m^\text{f})_i]_{i=1}^{N_m}$ and $\mathbf{E}_m^\text{a} = [(\mathbf{x}_m^\text{a})_i]_{i=1}^{N_m}$, respectively; here $(\mathbf{x}_m^\text{f})_i$ and $(\mathbf{x}_m^\text{a})_i$ denote the $i$th member in the forecast or analysis ensemble. We denote the means of the forecast and analysis ensemble by $\overline{\mathbf{x}}^\text{f}_m$ and $\overline{\mathbf{x}}^\text{a}_m$, respectively.

\subsection{Incorporation of model error}

\subsubsection{The model error}\label{sec:model_error}

\cite{narayan_sequential_2012} did not explicitly address model error covariances as part of the multi-model Kalman filter. \cite{yang_sequential_2017} did  include model errors in their multi-model filter equations, but did not discuss methods to estimate them. We stress here that considering model errors is critical for the MM-DA's correctly weighting models, and that the multi-model filter must therefore be supplemented by a model error estimation method.

We assume that the true state evolution of the system can be expressed, for each model $\mathbf{M}_m$, as
\begin{linenomath}
\begin{equation}
\mathbf{G}_m\mathbf{x}^\text{t}(t_{i}) = \mathbf{M}_{m}(t_{i-1})\mathbf{G}_m\mathbf{x}^\text{t}(t_{i-1}) + \bm{\eta}_m(t_{i-1}),
\end{equation}
\end{linenomath}
where $\mathbf{x}^\text{t}(t_i)$ is the true state at time $t_i$ and $\bm{\eta}_m$ is a model error with mean $\mathbf{0}$ and covariance $\mathbf{Q}_m$.

For model $m$, the forecast error covariance $\mathbf{P}_m^\text{f}$ at time $t_{i+1}$ can then be estimated by Eq.~\ref{eq:cov_prop}. This equation holds exactly only for a linear model \cite{tandeo_review_2020}. Thus $\mathbf{P}_m^\text{f}$ can be written as a sum of two terms,
\begin{linenomath}
\begin{equation}
\mathbf{P}_m^\text{f}(t_{i+1}) = \mathbf{P}_m^\text{p}(t_{i+1}) + \mathbf{Q}_m(t_{i}).\label{eq:pf}
\end{equation}
\end{linenomath}
The term $\mathbf{P}^\text{p}$ is sometimes called the predictability error \cite{berry_adaptive_2013}, and is due to the effect of the system's dynamics on the uncertainty in the initial conditions. Therefore Kalman filters, without incorporating $\mathbf{Q}$, are prone to underestimate $\mathbf{P}^\text{f}$.

Note that the assumption that the total forecast error can be decomposed as a sum of an initial-condition error and a model error becomes less justified at longer lead times, due to the correlations between the initial condition and model errors \cite{carrassi_model_2008,mitchell_accounting_2015}.

Besides the underestimation problem, the consideration of model error in MM-DA is critical, since estimating the forecast error covariance from the ensemble spread as in EnKFs may give similar weights to models of different accuracy. For example, in \cite{li_accounting_2009}, the perfect model was found to have similar spread to an imperfect model. Another issue is that a systematically overconfident model would be given higher weight if only the spread is accounted for.

Common ways to handle model error include: estimating the model error covariance matrix $\mathbf{Q}$ and using it to inflate the forecast covariance (additive inflation); inflating the forecast covariance with scalars (multiplicative inflation); or attempting to directly correct model error (bias correction). \cite{gharamti_enhanced_2018} discusses several additional methods.

Additive inflation generally works better than simple multiplicative inflation in accounting for model errors \cite{hamill_accounting_2005,li_accounting_2009,whitaker_evaluating_2012,raanes_extending_2015}, since the latter assumes that model errors will have the same structure as errors due to initial conditions, which is not generally the case. Estimating scalar inflation factors, however, is more feasible in high-dimensional and data-scarce settings than estimating the matrix  $\mathbf{Q}$. Moreover, there are methods for multiplicative covariance inflation that allow the inflation to vary in space and time \cite{anderson_spatially_2009,gharamti_enhanced_2018,tandeo_review_2020}. Such methods are likely to narrow the performance gap or surpass temporally fixed additive inflation.

Several sophisticated state-dependent bias correction schemes have been developed and used in DA \cite{li_accounting_2009,farchi_comparison_2021}. The best results are usually obtained by a combination of bias correction and inflation \cite{baek_local_2006,li_accounting_2009}, with the latter accounting for the model error remaining after the bias correction.

In this paper, we use additive inflation to account for model error. Future work could apply bias correction to each model in addition to inflation. In the algorithms that follow, we use $\mathbf{b}_m$ to refer to the bias of model $m$, when bias estimation is employed; otherwise, $\mathbf{b}_m = \mathbf{0}$.

\subsubsection{Estimation method and use in filtering}

In this paper, we use a simple, innovation-based estimation method for model error covariance, which we describe in \ref{sec:Q_est}. However, there are a variety of methods for estimating $\mathbf{Q}$, often simultaneously with estimating $\mathbf{R}$; see the reviews of  \cite{dunik_noise_2017} and \cite{ tandeo_review_2020}. When estimating $\mathbf{Q}$ is not computationally feasible, many methods for adaptive estimation of multiplicative covariance inflation are available, as described in the last subsection.

Several methods to estimate $\mathbf{Q}$, including the one we use, rely on the statistics of the innovations, i.e., of the differences between observations and forecasts. In order to compute innovations for our MM-EnKF, we must define an additional observation operator $\mathbf{H}_m: \mathbb{R}^{n_m}\to\mathbb{R}^p$ for each model, which maps the model space to the observation space. For the reference model $m=m_r$, $\mathbf{H}_{m_r} = \mathbf{H}$. When $\mathbf{G}_m$ is injective, $\mathbf{H}_m$ is given by
\begin{linenomath}
\begin{equation}
\mathbf{H}_m = \mathbf{G}_m^{\dagger}\mathbf{H}.\label{eq:hm}
\end{equation}
\end{linenomath}
In case $\mathbf{G}_m$ is not injective, $\mathbf{H}_m$ would have to be specified for every model. The innovations for model $m$ are given by $\mathbf{d}_m = \mathbf{y} - \mathbf{H}_m\mathbf{x}^\text{f}_m$.

Given an estimate of $\mathbf{Q}_m$, in order to account for it in the ensemble, samples drawn from the multivariate Gaussian distribution $\mathcal{N}(\mathbf{0}, \widetilde{\mathbf{Q}}_m)$ can be added to the $m^\mathrm{th}$ forecast ensemble \cite{mitchell_accounting_2015,asch_data_2016}, as done herein. \cite{mitchell_accounting_2015} found this stochastic method to perform better than directly inflating the covariance matrix. \cite{raanes_extending_2015} showed, however, that some methods work better for additive inflation in square-root filters than random sampling.

Here, we estimate $\mathbf{Q}_m$ for each model independently using the method described in \ref{sec:Q_est}. \cite{logutov_multi-model_2005}, though, showed that an error estimation method---in their case, the direct estimation of the forecast error covariance matrices $\mathbf{P}_m^\text{f}$---using all the models simultaneously can be more effective, especially when there is a small number of verifying observations. While not taken here, the latter approach could prove useful in the future.

\subsection{Ensemble perturbations}\label{ssec:perturb}

By applying the MM-DA framework directly to an EnKF, the iterative procedure results, prior to assimilating observations, in a combined multi-model forecast ensemble $\mathbf{E}_{1:M}^{\text{f}'}$. This ensemble lives in the reference model space, and has $N_{m_r}$ ensemble members. A disadvantage of this approach is that, already at the beginning of the forecast cycle, it reduces the number of ensemble members from $\sum_m N_m$ to $N_{m_r}$. Even though the information from these members is included in $\mathbf{E}_{1:M}^{\text{f}'}$, a larger ensemble helps  reduce sampling error.

Furthermore, once observations are assimilated, we obtain $\mathbf{E}^\text{a}$, an analysis ensemble in the reference model space. How does one then obtain the analysis ensemble $\mathbf{E}^\text{a}_m$ in each model space $m$, in order to use it as initial conditions for the next forecast cycle?

Previous work on MM-DA did not address these questions dealing with ensemble perturbations in an MM-EnKF. Here, we discuss three ways of doing so.

\begin{figure*}
\centering
\includegraphics[scale=1.2]{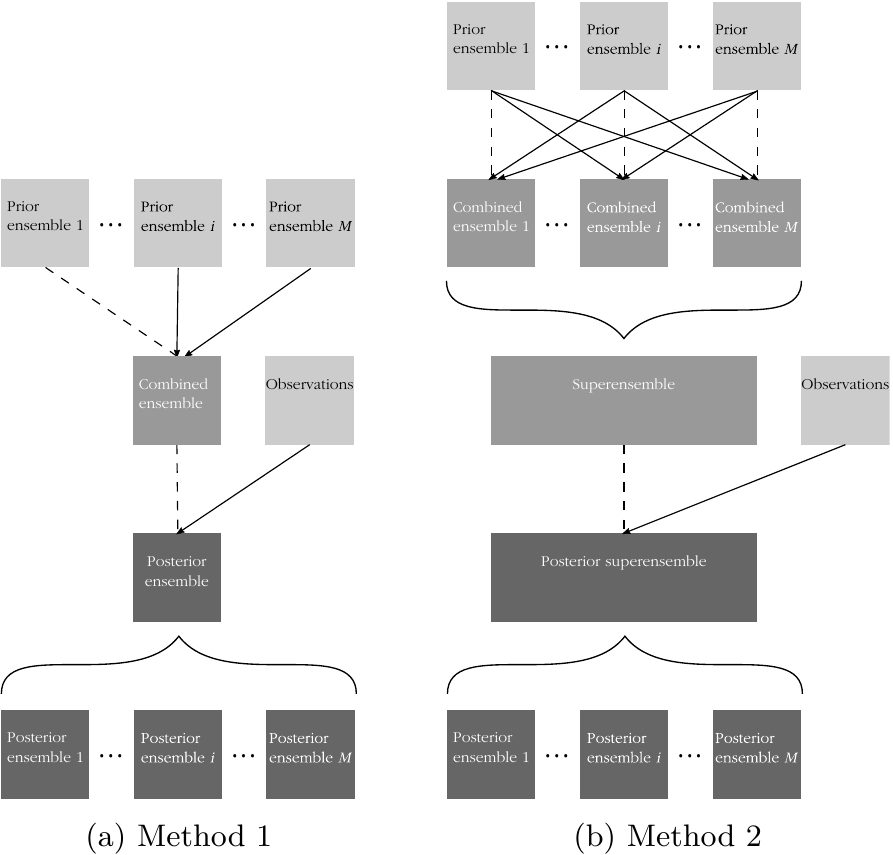}
\caption{Diagram of the proposed algorithms. In Method 1, one model is treated as the reference model. In Method 2, each model is treated as the reference model in turn, and the combined forecast ensembles are concatenated into a superensemble before assimilating observations. Arrows indicate information treated as observations in DA, dotted lines indicate those treated as background. Braces indicate concatenation for Method 2, and application of Eq.~\ref{eq:Ema} for Method 1.}
\label{fig:methods_panel}
\end{figure*}

\subsubsection{Method 1}

This method is a straightforward application of the iterative methodology described above: we simply compute $\mathbf{E}_{1:M}^{\text{f}'}$. After assimilating observations, we then take
\begin{linenomath}
\begin{equation}
\mathbf{E}^\text{a}_m = \mathbf{G}_m\mathbf{E}^\text{a}\label{eq:Ema}
\end{equation}
\end{linenomath}
as the analysis ensemble in model space $m$. A schematic diagram is shown in Fig.~\ref{fig:methods_panel}a.

This method has a disadvantage in terms of sampling error, as described above. Moreover, Eq.~\ref{eq:Ema} implies that each new model ensemble will now have the same number of ensemble members as that of the reference model: $N_m = N_{m_r}$. If $N_{m_r} \geq N_m$ for all $m\neq m_r$, though, a random choice of $N_m$ ensemble perturbations out of the $N_{m_r}$ could still be made for each $m\neq m_r$.

A related issue is that, for Method 1, each model's posterior ensemble has the same perturbations but transformed into the model space $m$, since $\mathbf{E}^\text{a} = \overline{\mathbf{x}}^\text{a} \mathbf{1}^T + (N-1)^{1/2}\mathbf{X}^\text{a}$ \cite{asch_data_2016} implies that $\mathbf{G}_m\mathbf{E}^\text{a} = \overline{\mathbf{x}}_m^\text{a} \mathbf{1}^T + (N-1)^{1/2}\mathbf{G}_m \mathbf{X}^\text{a}$, where $\mathbf{X}^\text{a}$ are the reference model's analysis ensemble perturbations and $\mathbf{1}$ is a vector of ones. This may reduce the effective number of ensemble members in the multi-model forecast ensemble, when the models are similar enough.

\subsubsection{Method 2}

We propose an alternative method for handling the ensemble perturbations. Here, we repeat the iterative procedure of Eq.~\ref{eq:models}  %\ref{eq:kalman_gain}--\ref{eq:kalman_pa} 
$m$ times at each assimilation step, changing the reference model to $m_r = m$ each time. Then, we have $m$ model ensembles, each in their own model space. We then map all these ensembles into a single model space, considering them as a single ``superensemble''. The observations are then assimilated into this superensemble, and the analysis ensemble members can be mapped back into their respective model spaces. A schematic diagram appears in Fig.~\ref{fig:methods_panel}b.

This method uses all the ensemble members in assimilating the observations, and will thus suffer from lower sampling error than Method 1. Furthermore, one obtains an analysis ensemble for each model which retains the number of ensemble members $N_m$, and has distinct ensemble perturbations for each model.

This method, though, has a larger computational cost than Method 1: the multi-model forecast combination cost in terms of operation count will increase by a factor of about $M$, although each of the $M$ assimilation steps can be done in parallel. Likewise, if the ensemble sizes are equal, the memory requirement for the analysis ensembles will increase by a factor of $M$. Furthermore, this method requires mappings from every model space $m$ to every other model space $m'$: $\mathcal{G}_{m\to m'}$. This is only possible if the mappings are invertible. Hence, this method is easiest to implement when all the models are in the same space or when there is a simple mapping between them, but it will not work with different dimensions of model space.

\subsubsection{Other approaches}

Lastly, the analysis ensemble could be regenerated for each $m\neq m_r$ by matching the known moments of the analysis distribution $\left(\overline{\mathbf{x}}_m^\text{a}, \mathbf{P}_m^\text{a}\right)$, as obtained from Eqs.~\ref{eq:xma} and \ref{eq:pma}. There is no unique set of ensemble members that possess these moments, but some that do can be generated by sampling from the multivariate normal $\mathcal{N}(\overline{\mathbf{x}}_m^\text{a}, \mathbf{P}_m^\text{a})$. 

Another way to generate appropriate ensemble members is to use sigma points as in the unscented Kalman filter \cite{julier_unscented_2004}. Doing so, however, requires at least $2n_m$ sigma points, which is not feasible for high-dimensional, computationally expensive models.

In the numerical experiments that follow, we compare only Methods 1 and 2.

\subsection{Computational considerations}\label{ssec:computation}

For the iterative form of MM-DA, the Kalman gain Eq.~\ref{eq:kalman_gain} can be written as
\begin{linenomath}
\begin{equation}
\mathbf{K}_m \left(\mathbf{G}_m\mathbf{P}_{1:m-1}^{\text{f}'}\mathbf{G}_m^T + \mathbf{P}_{m}^\text{f}\right) = \mathbf{P}_{1:m-1}^{\text{f}'} \mathbf{G}_m^T,\label{eq:gain_system}
\end{equation}
\end{linenomath}
where the linear system is solved for $\mathbf{K}_m$ in order to avoid explicit matrix inversion. Note that here, since the forecast of model $m$ is treated as an observation, we are required to solve a system in the model space $\mathbb{R}^{n_m}$. This can be too computationally expensive for high-dimensional models.

\subsubsection{Taking advantage of lower-dimensional models}

If only one of the models has very high dimension, that model can be chosen to be the reference model $m_r$. Then, in the iterative procedure, the inversions will only have to be done in the lower-dimensional model spaces, thus facilitating the computations.

Alternatively, if several models have high dimensions but only large-scale features are of interest, their forecasts could be mapped to a lower-dimensional space prior to assimilation, and the $\mathbf{G}_m$ modified accordingly. In the case of weather or climate models, this could consist in mapping the forecasts to a coarser grid. Possible solutions for multiple high-dimensional models with high-dimensional features that are relevant will be discussed below.

\subsubsection{Taking advantage of low rank}

When using an EnKF, both $\mathbf{P}_{1:m-1}^{\text{f}'}$ and $\mathbf{P}_{m}^\text{f}$ in Eq.~\ref{eq:gain_system} will be sample covariance matrices and they will be rank-deficient if the ensemble sizes are smaller than the model dimensions, as is typically the case. Localization generally increases the rank; when not applied, the low rank of these covariance matrices can be exploited to obtain a least-squares solution in $\mathcal{O}(n_m N_m'^2)$ operations, where $N_m'$ is the rank of the matrix $\mathbf{G}_m\mathbf{P}_{1:m-1}^{\text{f}'}\mathbf{G}_m^T + \mathbf{P}_{m}^\text{f}$ \cite{mandel_efficient_2006}.

\subsubsection{Right-multiplied ESRFs}

In the following approaches to efficient MM-EnKF implementation, an important role is played by square-root Kalman filters (SRFs) \cite{bellantoni_square_1967, bierman_factorization_1977}, and in particular ensemble SRFs (ESRFs) \cite{tippett_ensemble_2003}. In their historical account, \cite{grewal_applications_2010} state that the SRF is an ``improvement [...] over conventional Kalman filtering [achieving] `the same accuracy with half as many bits' of precision.''

An alternative form of the gain is obtained by applying the Sherman--Morrison--Woodbury formula \cite{hager_updating_1989} to Eq.~\ref{eq:gain_system}:
\begin{linenomath}
\begin{equation}
\mathbf{K}_m = ((\mathbf{P}_{1:m-1}^{\text{f}'})^{-1} + \mathbf{G}_m^T(\mathbf{P}_m^\text{f})^{-1}\mathbf{G}_m)^{-1}\mathbf{G}_m^T(\mathbf{P}_m^\text{f})^{-1}\label{eq:model_space}
\end{equation}
\end{linenomath}
Some ensemble Kalman filter variants use gains of the form Eq.~\ref{eq:model_space}, but express the analysis in the ensemble subspace \cite{asch_data_2016}. These are known as right-multiplied ESRFs \cite{sakov_relation_2011}. 

The ensemble transform Kalman filter \cite[ETKF:][]{bishop_adaptive_2001} is an important form of right-multiplied ESRF. The gain can be computed by  solving a linear system without explicitly inverting $\mathbf{P}_m^\text{f}$. If $\mathbf{P}_m^\text{f}$ is assumed to have a block-diagonal structure with relatively small blocks, the computation becomes feasible. This block-diagonal structure is intrinsic to analyses being done locally, as in the local ETKF \cite[LETKF:][]{hunt_efficient_2007}.

Similar to the low-rank case discussed above, when $\mathbf{P}_m^\text{f}$ is rank-deficient with rank $N_m$, its pseudoinverse can be computed in $\mathcal{O}(n_m N_m^2)$ operations.

\subsubsection{Structured covariance matrices}

Another way of making the matrix operations less expensive is to take either $(\mathbf{G}_m\mathbf{P}_{1:m-1}^{\text{f}'}\mathbf{G}_m^T + \mathbf{P}_{m}^\text{f})$ in gains of the form of Eq.~\ref{eq:gain_system} or $\mathbf{P}_m^\text{f}$ in gains of the form of Eq.~\ref{eq:model_space} to have a simplified structure. As discussed above, block-diagonal structure is one such possibility. Block-diagonality also enables the use of sequential EnKFs \cite{houtekamer_sequential_2001}. Several simplified structures were considered for observation error covariance matrices in \cite{stewart_data_2013}.

The simplest structure, but a rather restrictive one, is assuming the matrices to be diagonal; then the inverse is trivial to compute and store. The diagonality assumption is often made for the covariance matrices of observation errors. Note that if all the $\mathbf{P}^{\text{f}}_{m}$ are treated as diagonal in computing Eqs.~\ref{eq:mm_xa} and \ref{eq:mm_Pa} with the direct method, the solution corresponds to the minimum variance estimator when the weights for each $\mathbf{x}^\text{f}_m$ are vectors instead of matrices; see Corollary 2 in \cite{sun_multi-sensor_2004}.

For right-multiplied ESRFs, if we only impose the simplified structure when inverting $\mathbf{P}_m^\text{f}$, the simplified-structure assumption is not made for the forecast covariance of the reference model, $\mathbf{P}^{\text{f}}_{m_r}$.

In the experiments below, since the dimensionality is relatively low, we first apply localization to each $\mathbf{P}^{\text{f}}_{m}$ and then invert directly.

\subsection{Multi-model forecasting}\label{ssec:forecasting}

MM-DA can be used for real-time forecasting with multiple models by carrying out the iterative procedure for the available models and not assimilating any observations \cite{logutov_multi-model_2005, narayan_sequential_2012}. Doing so corresponds simply to the use of Eq.~\ref{eq:models} to combine the multiple models.

We can let $\mathbf{R}^{-1} \to \mathbf{0}$ in Eq.~\ref{eq:objective}, since this limit of infinite variance simply corresponds to no observations being available. Then, the Eqs.~\ref{eq:mm_xa} and \ref{eq:mm_Pa} of MM-DA for the analysis state and covariance become
\begin{linenomath}
\begin{subequations}
\begin{align}
\mathbf{x}^\text{a} &= \mathbf{P}^\text{a}\left(\sum_{m=1}^M \mathbf{G}_m^T \left(\mathbf{P}^\text{f}_m\right)^{-1}\mathbf{x}^\text{f}_m \right),\label{eq:mm_forecasts}\\
\mathbf{P}^\text{a} &= \left(\sum_{m=1}^M \mathbf{G}_m^T \left(\mathbf{P}^\text{f}_m\right)^{-1}\mathbf{G}_m\right)^{-1}.
\end{align}
\end{subequations}
\end{linenomath}
Thus, MM-DA neatly handles multi-model forecasting in addition to DA. Note that, when $\mathbf{G}_1 = \mathbf{G}_2 = \cdots = \mathbf{G}_M = \mathbf{I}$ and $\mathbf{P}^\text{f}_1 = \mathbf{P}^\text{f}_2 = \cdots = \mathbf{P}^\text{f}_M$, Eq.~\ref{eq:mm_forecasts} reduces simply to the unweighted multi-model average.

When forecasting at long lead times, it can be helpful to apply MM-DA recursively at intermediate leads. The set of model error covariance matrices should be specified for each lead time; it is known in the seasonal climate prediction context, for example, that the ``best model'' can depend on the lead time \cite{hagedorn_rationale_2005}.

Assume that we have estimated the model error covariance matrix for each model $m$ at different intermediate lead times $k\tau$, which we denote by $\mathbf{Q}_m^{k \tau}$. Then, if the desired forecast horizon is $T = K \tau$, MM-DA can be applied first at lead time $\tau$ with model error covariance matrices $\mathbf{Q}_m^\tau$. The analysis for this horizon is then used as an initial condition for the forecasts out to time $2\tau$, whereupon MM-DA is applied with $\mathbf{Q}_m^{2\tau}$, etc. This recursive method tends to perform better than directly applying MM-DA at horizon $T$, since the trajectory is repeatedly corrected.

One may wonder whether, at long lead times, when the error growth of a nonlinear forecast model ceases to obey linearized dynamics, Eq.~\ref{eq:pf}  for the forecast error covariance is still a good approximation. Here, it is more useful to think of $\mathbf{Q}$ as an additive inflation that compensates for overconfidence in the prediction.

\subsection{Filter algorithm}

Localization is critical for EnKFs \cite{carrassi_data_2018}. Here, we apply localization at each step of the iterative procedure, and also when observations are assimilated.

We use the left-multiplied form of the ESRF, as described in \cite{sakov_relation_2011}, for both the multi-model combination and the assimilation of observations. This EnKF is a deterministic filter for which it is particularly simple to express covariance localization.

The left-multiplied ESRF equations are given by
\begin{linenomath}
\begin{subequations}
\begin{align}
& \mathbf{X} = {(N - 1)^{-1/2}} (\mathbf{E}^\text{f} - \overline{\mathbf{x}}^\text{f} \mathbf{1}^T) ,\\
& \mathbf{P}^\text{f} = \bm{\rho}\circ(\mathbf{X}\mathbf{X}^T),\\
& \mathbf{K} = \mathbf{P}^\text{f}\mathbf{H}^T (\mathbf{H}\mathbf{P}^\text{f}\mathbf{H}^T + \mathbf{R})^{-1},\\
& \overline{\mathbf{x}}^\text{a} = \overline{\mathbf{x}}^\text{f} + \mathbf{K}(\mathbf{y} - \mathbf{H}\overline{\mathbf{x}}^\text{f}),\\
& \mathbf{E}^\text{a} = \overline{\mathbf{x}}^\text{a} \mathbf{1}^T + (N - 1)^{1/2}(\mathbf{I} - \mathbf{K}\mathbf{H})^{1/2}\mathbf{X},
\end{align}
\end{subequations}
\end{linenomath}
where $\bm{\rho}$ is the localization matrix; $\circ$ is the Hadamard, or element-wise, product; and $\mathbf{X}$ are the normalized ensemble perturbations.

In the iterative procedure, we use the ensemble mean $\overline{\mathbf{x}}_m^\text{f}$ of model $m$ as the observation for the multi-model ensemble $\mathbf{E}_{1:m-1}^{\text{f}'}$.

This ESRF form is not efficient for high-dimensional systems, since the update is done in the state space instead of the ensemble space. For high-dimensional systems, right-multiplied ESRFs are more practical. See section~\ref{ssec:computation} for more details on computational issues.

\subsection{Inflation}\label{ssec:inflation}

EnKFs generally underestimate the forecast covariance due to model and sampling errors, thus imposing the need for inflation \cite{carrassi_data_2018}. While we attempted to account for the model error in each individual model, we found that the multi-model forecast covariance is usually still underestimated, and the underestimation increases with $M$.

This underestimation is due to the assumption that the models are unbiased, and that the errors for distinct models are independent of one another. That is, if the model forecasts were unbiased and independent, one would expect the error in a multi-model average to decrease as $M^{-1/2}$, but this does not happen. See \cite{knutti_challenges_2010} and \cite{christiansen_understanding_2020} for an explanation of this phenomenon in multi-model ensembles. Furthermore, in assimilating forecast states of one model into another one, which has a different attractor, one inherently encounters representation error \cite{hodyss_error_2015}. Hence, we also need to apply inflation to the multi-model forecast.

Here, we use a simple multiplicative covariance inflation scheme, with the inflation factor $\hat{\lambda}$ estimated as in \cite{tandeo_review_2020}:
\begin{linenomath}
\begin{equation}
\hat{\lambda} = \frac{\mathbf{d}^T\mathbf{d} - \operatorname{tr}(\mathbf{R})}{\operatorname{tr}(\mathbf{H}\mathbf{P}^\text{f}\mathbf{H}^T)}.\label{eq:sim_inflation}
\end{equation}
\end{linenomath}
Since the inflation is applied to the multi-model forecast, we take $\mathbf{d} = \mathbf{y} - \mathbf{H}\mathbf{x}_{1:M}^{\text{f}'}$ and $\mathbf{P}^\text{f} = \mathbf{P}_{1:M}^{\text{f}'}$ and then apply a temporal smoothing, as in Eq.~\ref{eq:mean}, which yields
\begin{linenomath}
\begin{equation}
\widetilde{\lambda}(k+1) = \gamma \hat{\lambda}(k) + (1 - \gamma)\widetilde{\lambda}(k),
\end{equation}
\end{linenomath}
for some $0<\gamma<1$. Note that the numerator of Eq.~\ref{eq:sim_inflation} is not guaranteed to be positive, although its expected value is. However, negativity of $\hat{\lambda}$ does not pose a problem as long as the smoothed estimate $\widetilde{\lambda}$ is positive. Encountering a negative $\widetilde{\lambda}$ suggests either a misspecification of the error covariance matrices or a $\gamma$-value that is too large, allowing for rapid fluctuations in $\widetilde{\lambda}$.

Due to the MM-DA--specific reasons above, the resulting values of $\lambda$ are higher than typically encountered with regular covariance inflation: in the experiments below, for instance, we have encountered $\widetilde{\lambda}$-values as large as 4.

\subsection{Algorithms}
We are ready now to summarize in pseudo-code the two proposed versions of the MM-EnKF, as Algorithms~\ref{alg:method1} and \ref{alg:method2}. To maintain generality, we define the following \verb|DA_step| function, which represents the analysis step for any EnKF, and in which the observation operator $\mathcal{H}$ is kept as possibly nonlinear, since ensemble Kalman filters allow for nonlinear observation operators:

\LinesNumberedHidden
\RestyleAlgo{boxed}
\begin{algorithm}[h]
\SetKwProg{Fn}{function}{}{}
\SetKwFunction{FDA}{DA\_step}
\Fn{\FDA}

\KwIn{}
\begin{itemize}
  \setlength\itemsep{-0.3em}
\item $\mathbf{E}^\mathrm{f}$, the prior ensemble
\item $\mathbf{y}$, the observation vector
\item $\mathbf{R}$, the observation error covariance
\item $\mathcal{H}$, the observation operator
\end{itemize}\vspace{-0.7em}
\KwOut{$\mathbf{E}^\mathrm{a}$, the posterior ensemble}    
\end{algorithm}

\LinesNumbered
\RestyleAlgo{ruled}
\begin{algorithm}[h]
\caption{Multi-model ensemble Kalman filter step (Method 1)}\label{alg:method1}
\SetKwFunction{FDA}{DA\_step}
\SetKw{KwIn}{in}
\SetAlgoLined
  \tcc{Inflate ensemble members from estimated model error distribution}
% \For{$t$ = 0:$\Delta t$:$T$}{
  \For{$m$ \KwIn $(1,\ldots,M)$}{\label{alg:Q_begin}
  %$\mathbf{Q}(t+\Delta t) = \rho\widetilde{\mathbf{Q}}(t) + (1 - \rho)\mathbf{Q}(t)$

  %estimate $\mathbf{Q}_m$% from innovations
  
  \For{$i$ \KwIn $(1,\ldots,N_m)$}{
  $\bm{\eta}_i \sim \mathcal{N}(-\mathbf{b}_m, \mathbf{Q}_m)$

  $\mathbf{x}_i^\mathrm{f} = \mathbf{x}_i^\mathrm{f} + \bm{\eta}_i$
  }
  }\label{alg:Q_end}

  \tcc{Assimilate the other model forecasts into the reference model ensemble}
	
	$\mathbf{E}^{\text{f}'}_{1:1} = \mathbf{E}_{1}^\mathrm{f}$
	
	\For{$\ell$ \KwIn $(2, \ldots, M)$}{
	$\mathbf{E}^{\text{f}'}_{1:\ell}$=\FDA{$\mathbf{E}^{\mathrm{f}'}_{1:\ell-1}$,
	$\overline{\mathbf{x}}^\mathrm{f}_{\ell}$,
	$\mathbf{P}_{\ell}^\mathrm{f}$, $\mathcal{G}_{\ell}$}
	}
  
  $\mathbf{E}^\mathrm{f} = \overline{\mathbf{x}} + \lambda^{1/2} (\mathbf{E}_{1:M}^{\mathrm{f}'} - \overline{\mathbf{x}})$

  \tcc{Assimilate observations}
  $\mathbf{E}^\text{a}$ = \FDA{$\mathbf{E}^\mathrm{f}$, $\mathbf{y}$, $\mathbf{R}$, $\mathcal{H}$}\label{alg:assimilate_obs_1}

  \tcc{Integrate each posterior ensemble to the next time}
  \For{$m$ \KwIn $(1,\ldots,M)$}{\label{alg:int_begin}
  
  $\mathbf{E}^\text{a}_m = \mathcal{G}_m((\mathbf{E}^\text{a})_m)$
  
  %obtain ICs as $\mathbf{E}^\text{a}_m = \mathbf{H}_{m_r\to 1}\mathbf{E}^\text{a}$

  $\mathbf{E}^\mathrm{f}_m(t+\Delta t) = \mathcal{M}_{t\to t+\Delta t}(\mathbf{E}^\text{a}_m(t))$
  }\label{alg:int_end}
% }
\end{algorithm}

\LinesNotNumbered
\RestyleAlgo{ruled}
\begin{algorithm}[h]
\caption{Multi-model ensemble Kalman filter step (Method 2)}\label{alg:method2}
\SetKwFunction{FDA}{DA\_step}
\SetKw{KwIn}{in}
\SetAlgoLined
  \tcc{Inflate ensemble members from estimated model error distribution}
  \nlset{1--6}(Same as lines \ref{alg:Q_begin}--\ref{alg:Q_end} in Algorithm 1)
\LinesNumbered\setcounter{AlgoLine}{6}

  \tcc{For each model ensemble, assimilate the other model forecasts into it}
  \For{$m$ \KwIn $(1, \ldots, M)$}{
  	  $O = (m, 1, \ldots, m - 1, m + 1,\ldots, M)$ \tcp{Order of assimilation}

	  $\mathbf{E}^{\text{f}'}_{m,1:1} = \mathbf{E}_{m}^\mathrm{f}$

	  \For{$\ell$ \KwIn $(2, \ldots, M)$}{
	  $\mathbf{E}^{\text{f}'}_{m,1:\ell}$=\FDA{$\mathbf{E}^{\mathrm{f}'}_{m,1:\ell-1}$,
	  $\overline{\mathbf{x}}^\mathrm{f}_{O_\ell}$,
	  $\mathbf{P}_{O_\ell}^\mathrm{f}$, $\mathcal{G}_{m\to O_\ell}$}

	  }
  }
  \tcc{Form a "superensemble" from all the ensembles}
  $\mathbf{E}^\mathrm{f} = [\mathcal{G}_{1\to m_r}(\mathbf{E}^{\text{f}'}_{1,1:M})\cdots \mathcal{G}_{M\to m_r}(\mathbf{E}^{\text{f}'}_{M,1:M})]$
  
  $\mathbf{E}^\mathrm{f} = \overline{\mathbf{x}} + \lambda^{1/2} (\mathbf{E}^\mathrm{f} - \overline{\mathbf{x}})$

  \tcc{Assimilate observations}
  $\mathbf{E}^\text{a}$ = \FDA{$\mathbf{E}^\mathrm{f}$, $\mathbf{y}$, $\mathbf{R}$, $\mathcal{H}$}\label{alg:assimilate_obs_2}
\LinesNotNumbered

  \tcc{Integrate each posterior ensemble to the next time}
 \nlset{17--20}(Same as lines \ref{alg:int_begin}--\ref{alg:int_end} in Algorithm 1)
% }
\end{algorithm}

Note that, for real-time forecasting, line~\ref{alg:assimilate_obs_1} in the pseudocode for Method 1 or line~\ref{alg:assimilate_obs_2} in the pseudocode for Method 2 is removed.

\section{Relation to other methods}

\subsection{MM-EnKF properties}

The MM-EnKF has the following properties, compared to other methods for multi-model DA and forecasting:
\begin{itemize}
\item The method is a natural generalization of the standard Kalman filter to multiple models, and can be derived from both the variational and Bayesian viewpoints \cite{narayan_sequential_2012}, as well as from linear minimum variance estimation (see \ref{sec:blue_proof}). This fact allows for the use of well-understood DA methods, and the theoretical apparatus of optimal state estimation and Kalman filters \cite[e.g.,][]{jazwinski_stochastic_1970,simon_optimal_2006}.

\item The methods reviewed in section \ref{sec:intro} mostly involve scalar weights. Here, the weights are matrices, which allows for variables to be weighted differently. In the case of spatiotemporal models, this allows the weights assigned to each model to vary in space; this is important in the case of atmospheric models, where model skill can be highly spatially inhomogeneous \cite{du_multi-model_2017}.% Moreover, unlike multi-model superensembles \cite{krishnamurti_review_2016}, the weights for one variable can affect others through cross-covariances in the forecast covariance matrices.

\item Each model can have its own model space. Most of the other reviewed methods do not allow for this, instead assuming a common model space. Distinct model spaces allow for the combination of models of different resolutions, those that predict different variables, or those that are restricted to different spatial domains. Some examples of such scenarios are provided in section \ref{sec:experiments}.

\item A common problem of adaptive multi-model methods is the weight of useful models converging to 0 \cite{smith_designing_2020}. With MM-DA, this problem does not occur as long as the filter is stable, since this would require $(\mathbf{P}^\text{f}_m)^{-1}\to \mathbf{0}$. This feature may have its  downside when a model is consistently detrimental.

\item If all models are biased in one direction, Bayesian model averaging will result in a forecast worse than the best model. This is not the case with multi-model DA \cite{narayan_sequential_2012}.

\item The MM-EnKF methodology provides probabilistic analyses and forecasts, using ensembles. Many of the methods for forecast combination reviewed herein assume a single deterministic forecast for each model, and do not account for uncertainty.

\item DA is designed for forecast problems, and MM-DA is shown in section \ref{sec:experiments} to improve forecast skill. However, some multi-model methods target instead improving climatology, i.e., the system's long-term statistics. In MM-DA, the $\mathbf{Q}_m$ are specified for a specific lead time; it is not clear how---or whether---these $\mathbf{Q}_m$'s can be adequately adapted to capture climatological error instead. It is often the case, though, that long-term systematic errors are similar to those at short timescales \cite{rodwell_using_2007,martin_analysis_2010}.

\item Several authors  \cite{ojeda_adaptive_2013,du_multi-model_2017,chen_multi-model_2019,bach_ensemble_2021,chattopadhyay_towards_2022,potthast_data_2022,counillon_framework_2022} explored the assimilation of forecasts as pseudo-observations. In many of these works, however, the error covariance assigned to the pseudo-observations was not defined in a consistent way, or the generalization to more than two models was not clear. MM-DA also assimilates forecasts as if they were observations, but in a consistent mathematical framework.

\item \cite{rainwater_mixed-resolution_2013} formulated an EnKF that uses ensembles at two different resolutions to compute the background covariance matrix, with a parameter that sets the weights given to each one. The low-resolution state forecast was not used. \cite{hoel_multilevel_2016,hoel_multilevel_2020} combined forecasts at different resolutions in an EnKF, but they did not weight them differently. \cite{popov_multifidelity_2021} combined models of different fidelities in an EnKF with a control variate approach. In section \ref{sec:experiments}, we will show how the MM-EnKF can effectively incorporate forecasts at different resolutions and fidelities.

\item In the terminology of \cite{mallet_ozone_2009}, MM-DA is a convex sequential aggregation rule.
\end{itemize}

\subsection{Connection to synchronization}

To combine forecasts of two models we have, from Eqs.~\ref{eq:kalman_fcst} and \ref{eq:kalman_gain},
\begin{linenomath}
\begin{subequations}
\begin{align}
\mathbf{x}_1 &= \mathbf{x}_{1}^\text{f} + \mathbf{K}_2(\mathbf{x}_2^\text{f} - \mathbf{G}_{1\to 2}\mathbf{x}_{1}^\text{f}),\\
\mathbf{x}_2 &= \mathbf{x}_{2}^\text{f} + \mathbf{K}_1(\mathbf{x}_1^\text{f} - \mathbf{G}_{2\to 1}\mathbf{x}_{2}^\text{f}),
\end{align}
\label{eq:synch}
\end{subequations}
\end{linenomath}
where
\begin{linenomath}
\begin{subequations}
\begin{align}
\mathbf{K}_1 &= \mathbf{P}_2^\text{f}\mathbf{G}_{2\to 1}^T(\mathbf{G}_{2\to 1}\mathbf{P}_2^\text{f}\mathbf{G}_{2\to 1}^T + \mathbf{P}_1^\text{f})^{-1},\\
\mathbf{K}_2 &= \mathbf{P}_1^\text{f}\mathbf{G}_{1\to 2}^T(\mathbf{G}_{1\to 2}\mathbf{P}_1^\text{f}\mathbf{G}_{1\to 2}^T + \mathbf{P}_2^\text{f})^{-1}.
\end{align}
\end{subequations}
\end{linenomath}
Here, $\mathbf{G}_{1\to 2}$ is the matrix mapping from a state in model space 1 to the corresponding state in model space 2, and vice-versa for $\mathbf{G}_{2\to 1}$. Thus, each model is being nudged towards the forecast of the other. This mutual nudging connects MM-DA to the synchronization view of DA \cite{carrassi_data_2008, abarbanel_unifying_2017, penny_mathematical_2017, penny_strongly_2019}: the multi-model combination step can be considered a form of impulsive synchronization between the models.

In the connected supermodelling approach of \cite{selten_simulating_2017}, connection terms between model states are introduced into the model equations. The connection coefficients are gathered into matrices $\mathbf{C}$, which can be identified with the gain matrices $\mathbf{K}_i$ in Eq.~\ref{eq:synch}. We note, however, that MM-DA differs from the approach of \cite{selten_simulating_2017}, as the latter directly estimates the connection coefficients by minimizing a cost function with training data. Additionally, the supermodelling approach uses static and diagonal $\mathbf{C}$, does not allow for different model spaces, and does not consider ensembles of each model. Future work could compare the connection coefficients obtained by connected supermodelling with the gains $\mathbf{K}_i$ obtained by MM-DA. Since supermodels are typically formulated in continuous time, determining the exact relationship between MM-DA and supermodels necessitates the derivation of the continuous-time analogue of the multi-model Kalman filter, namely a multi-model Kalman--Bucy filter.

A similar connection can be made between MM-DA and weighted supermodelling: in the latter, the supermodel tendency is a weighted average of the individual model tendencies \cite{schevenhoven_improving_2019}, while in MM-DA the analysis is a weighted average of the model forecasts (Eq.~\ref{eq:mm_forecasts}). \cite{wiegerinck_limit_2013} showed that a connected supermodel becomes a weighted supermodel in the limit of large couplings.

\section{Numerical experiments}\label{sec:experiments}
%\begin{table}
%	\begin{tabular}{llllll}
%		Scenario & Cycles & Spin-up cycles & $\Delta t$ & $\delta$ & $\gamma$ \\
%		\hline
%		1 & 5~000 & 3~000 & 0.05 & $10^{-3}$ & $10^{-2}$ \\
%		LR & & $10^{-3}$ & $10^{-2}$ \\
%		MM-EnKF & & $10^{-3}$ & $10^{-2}$ \\
%	\end{tabular}
%	\caption{Parameters for each experiment}
%	\label{table:experiment_parameters}
%\end{table}

Previously, MM-DA was only tested on very low-dimensional models with non-chaotic behavior \cite{narayan_sequential_2012, yang_sequential_2017}, and recursive multi-step forecasts were not tested. Methods for multi-model forecasting have often been tested with perfect observations for calibration, single forecasts for each model rather than ensembles, and models that all share the same space \cite{schevenhoven_efficient_2017,schevenhoven_improving_2019}; several papers, though, have extended this work to noisy observations \cite{du_multi-model_2017,schevenhoven_training_2022}. Here, we conduct twin experiments of the proposed method for both DA and forecasting in various settings, including models of different dimensionality and different-sized ensembles. Noisy observations are used for the model error estimation in all cases.

\subsection{Experimental set-up}\label{ssec:setup}

In the following numerical experiments, we use the Lorenz96 \cite{lorenz_predictability:_1996} model, except that we allow a different forcing $F_i$ for each site:
\begin{linenomath}
\begin{equation}
\frac{\text{d}x_i}{\text{d}t} = -x_{i-1}(x_{i-2} + x_{i+1}) - x_i + F_i;
\end{equation}
\end{linenomath}
here the indices $i$ range from 1 to $D$ and are cyclical. We use $D = 40$ variables in the experiments that follow.

The true model here has $F_i = 8$ for $1 \leq i \leq 10$, $F_i = 10$ for $11 \leq i \leq 20$, $F_i = 12$ for $21 \leq i \leq 30$, and $F_i = 14$ for $31 \leq i \leq 40$, similar to \cite{du_multi-model_2017}. We then define four imperfect forecast models to be used in the experiments, having fixed $F \equiv 8$, 10, 12, and 14 for all $i$.

We also use the two-scale version of model \cite{lorenz_predictability:_1996}:
\begin{linenomath}
\begin{subequations}\label{eq:Lor96}
\begin{align}
\frac{\text{d}x_i}{\text{d}t} &= -x_{i-1}(x_{i-2} + x_{i+1}) - x_i + F_i - \frac{hc}{b}\sum_{j=1}^n y_{j,i}, \label{eq:LS}\\
\frac{\text{d}y_{j, i}}{\text{d}t} &= -cby_{j+1,i}(y_{j+2,i} - y_{j-1,i}) - cy_{j,i} + \frac{hc}{b}x_i,  \label{eq:SS}
\end{align}
\end{subequations}
\end{linenomath}
where the indices $i$ range from 1 to $D$, the indices $j$ range from 1 to $d$, $y_{d+1,i} = y_{1,i+1}$, and $y_{0,i} = y_{d,i-1}$. The $y_{j_i}$ variables represent smaller-scale dynamics, which interact with the larger-scale $x_i$'s. We set $D=20$, $d=10$, $h=1$, $b=10$, $c=10$. With these parameters, the timescale is about 10 times as fast for the $y_{j_i}$'s as for the $x_i$'s. Given the full state vector containing both the $x$ and $y$ variables,
\begin{equation}
	\text{vec}\begin{pmatrix}
		x_1 & y_{1, 1} & y_{2, 1} & \cdots & y_{d, 1}\\
		x_2 & y_{1, 2} & y_{2, 2} & \cdots & y_{d, 2}\\
		\vdots & \vdots & \vdots & \ddots & \vdots\\
		x_D & y_{1, D} & y_{2, D} & \cdots & y_{d, D}
	\end{pmatrix},
\end{equation}
where $\operatorname{vec}$ is the vectorization operator which stacks the columns of the matrix on top of one another to obtain a column vector, the corresponding $\mathbf{G}_2$ is the $(d+1)D \times D$ matrix
\begin{equation}
	(\mathbf{G}_2)_{i,j} = \begin{cases}
		1, & \text{if } i = 1\\
		0, & \text{otherwise.}
	\end{cases}
\end{equation}

The time integrations used the fourth-order Runge--Kutta scheme. For the single-scale Lorenz96 model, we use a timestep of $\Delta t=0.05$, and for the two-scale one we use $\Delta t=0.005$.

For localization, we use the Gaspari--Cohn correlation function \cite{gaspari_construction_1999}. For experiments with the single-scale model, we use a localization radius of 4. For experiments with the two-scale model, we apply a localization radius of 4 to the $x_i$ variables and a radius of 40 to the $y_{j,i}$ ones. Cross-scale interactions are localized such that observations of the $y_{j,i}$ are allowed to influence the corresponding $x_i$, and vice versa.

In the following experiments, we compare the MM-EnKF to the unweighted multi-model ensemble (MME), wherein the multiple single-model ensembles are treated identically as a single ensemble, except that each model ensemble is inflated using the appropriate $\mathbf{Q}_m$. We also compare the results to each of the individual single-model ensembles, again inflated by their respective $\mathbf{Q}_m$. Moreover, scalar inflation is applied for both the MME and the MM-EnKF, as described in section~\ref{ssec:inflation}.

\subsection{Experiments with parametric model error}
\subsubsection{Multi-model DA}\label{ssec:MM_DA}
We test out Methods 1 and 2 of the MM-EnKF with the four imperfect models. We use an analysis window, or time interval over which observations are assimilated, of 0.2; $\mathbf{R} = 0.25 \mathbf{I}_{40}$; $\delta = 10^{-3}$ for the model error estimation; and $\gamma = 10^{-2}$ for the inflation estimation. Here we fully observe the state, but test partial observations in section \ref{ssec:partial}.

\begin{figure*}%[hb]
	\centering
	\includegraphics[scale=1.0]{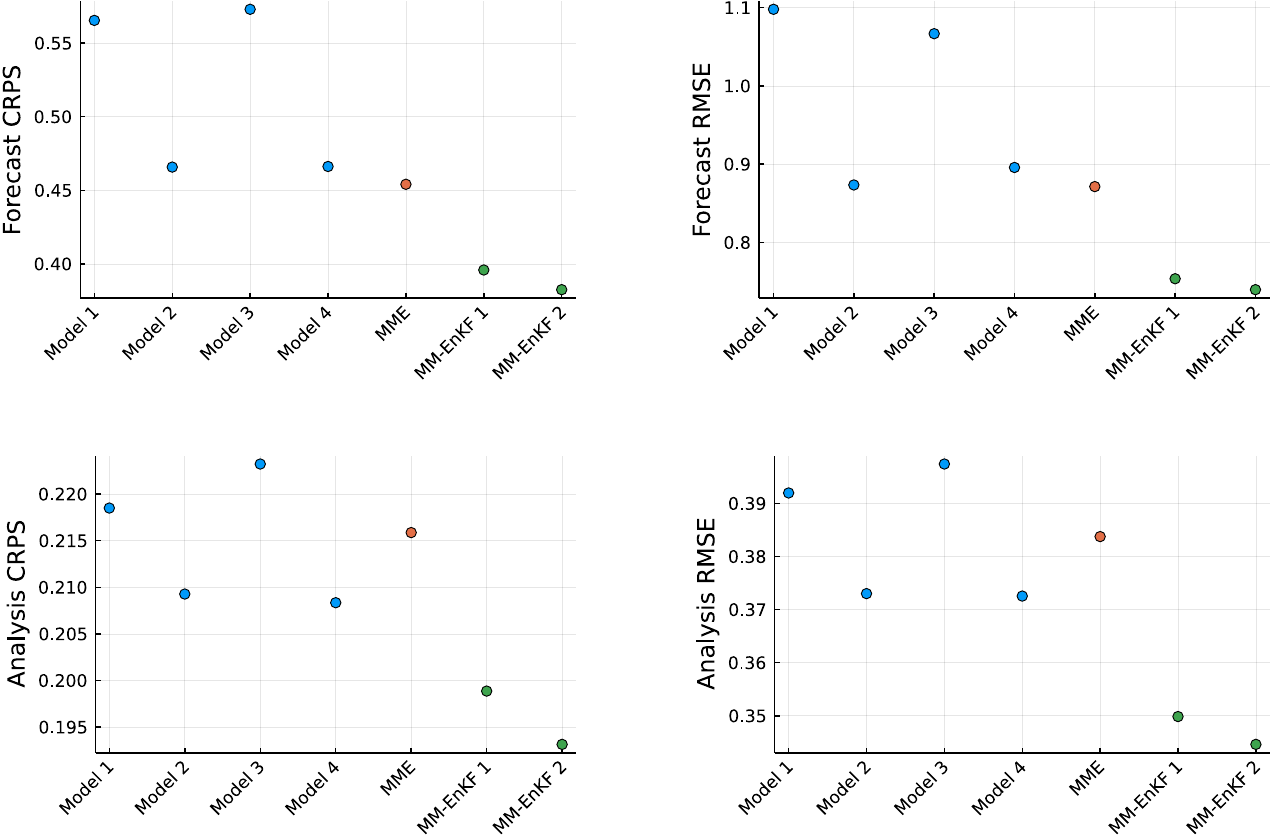}
	\caption{Overall performance of the MM-EnKF, in terms of both root-mean-square error (RMSE) and the continuous ranked probability score (CRPS). Here and in the subsequent experiments, we use suitably defined versions of the Lorenz96 model \protect\cite{lorenz_predictability:_1996}. The error bars are too small to be visible and hence none are plotted.}
	\label{fig:mmda}
\end{figure*}

For the MME and the MM-EnKF, we use 20 ensemble members for each model. In order to have a fair comparison, 80 ensemble members are used for each of the individual model experiments. We carry out 10~000 DA cycles, and average the error over the last 2~000.

Figure~\ref{fig:mmda} shows the results for the forecast and analysis errors. The forecast errors are for forecasts initialized from the analyses obtained by the filter, for a lead time equal to the analysis window. In addition to the root-mean-square error (RMSE), we use the continuous ranked probability score \cite[CRPS:][]{hersbach_decomposition_2000}, a probabilistic error metric, to measure the discrepancy between the ensemble and the true probability distributions. We apply the univariate CRPS along each dimension, and then take the mean. A strength of the CRPS is that it is a strictly proper scoring rule \cite{wilks_statistical_2019}.

The regular MME performs slightly better than the best model in terms of forecast error, and worse than the best model in terms of analysis error. The MM-EnKF, though, performs better than the MME and any individual model, in both forecast and analysis errors, and Method 2 has a slight edge over Method 1. The latter fact is likely due to Method 2 using a larger ensemble than Method 1 when assimilating the observations; see section \ref{ssec:perturb}.

\paragraph{Impact of model error estimation}

\begin{figure*}
	\centering
	\includegraphics{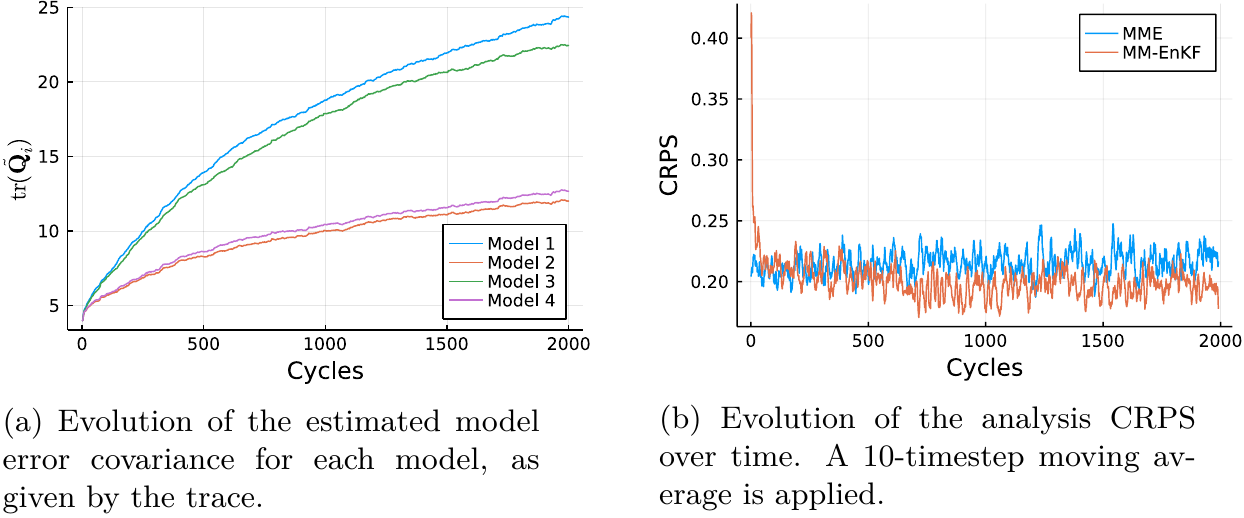}
	\caption{(a) Evolution of estimated model error covariance for the four models. Simultaneously in time, (b) shows a comparison of the evolution of the CRPS error metric for the unweighted MME and the MM-EnKF.}
	\label{fig:convergence}
\end{figure*}

To see the effect of model error estimation on the performance of the MM-EnKF methods, we run it simultaneously with the DA itself, and consider the time evolution of the weights and analysis error. In Fig.~\ref{fig:convergence}a we see the model weights evolving: initially assigned the same model error covariance, the model error estimation procedure estimates a higher error for models 1 and 3, and they are thus weighted less in the DA. Note that we show here only the trace, but in reality the weights are not the same for all variables.

In Fig.~\ref{fig:convergence}b, the analysis error is shown over the same time interval. Initially, with the same weight for each model, the MM-EnKF performs worse than an unweighted MME. However, as the model error estimation becomes more accurate, the MM-EnKF reaches a lower asymptotic error than the MME.

\paragraph{Impact of assimilation order}

\begin{figure}
	\centering
	\noindent\includegraphics[scale=0.3]{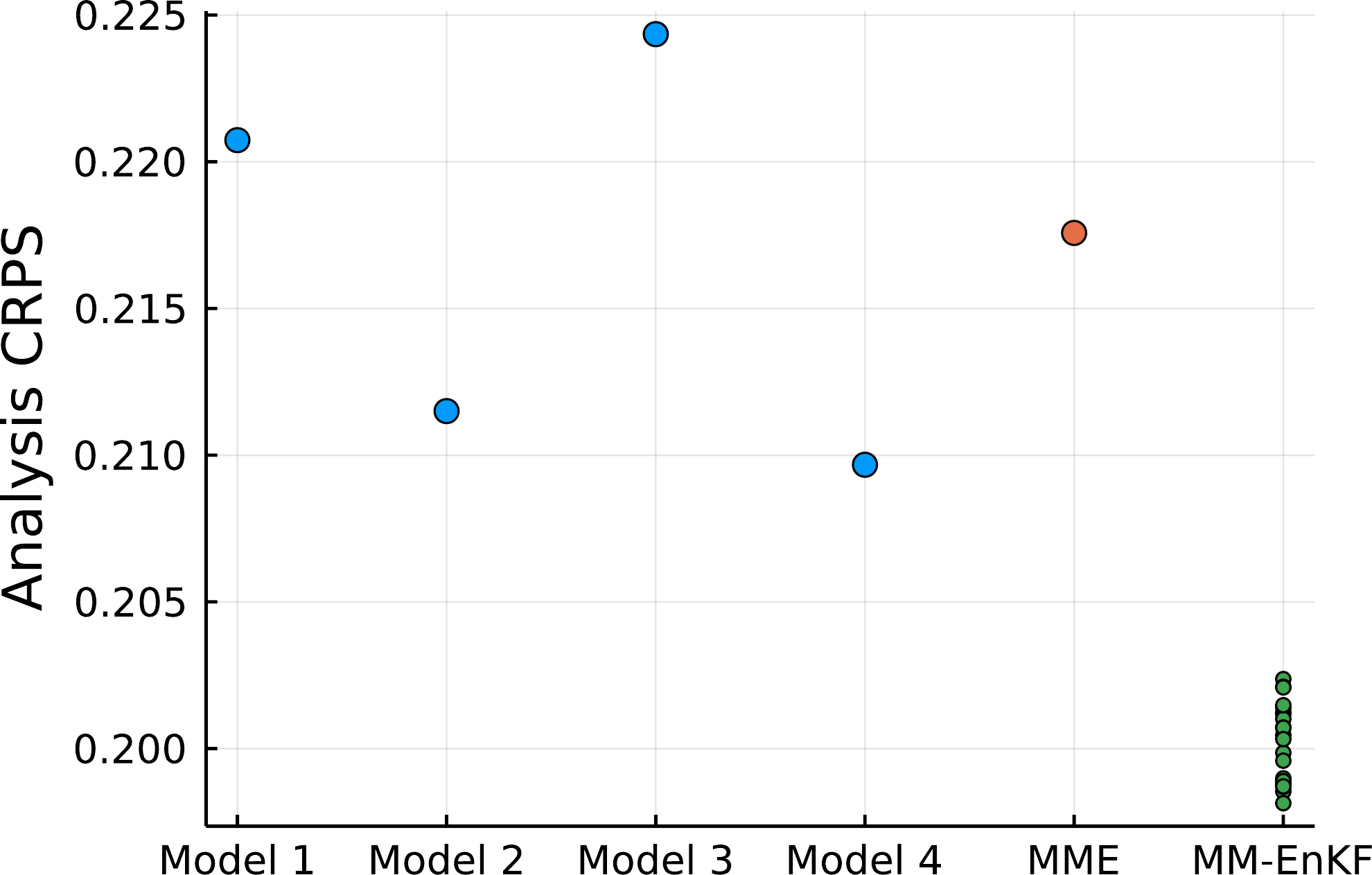}
	\caption{Performance of Method 1 of the MM-EnKF, with all permutations in the order of assimilating the four models plotted in green. The single-model analyses and unweighted MME are included for reference.}
	\label{fig:order}
\end{figure}

To test the effect of the order in which the models are assimilated, we repeated the experiment with the 24 = 4! possible permutations of the model orders. The results are shown in Fig.~\ref{fig:order}. In this case, model order is not very significant, and all the orders result in errors smaller than the best model and the MME. Furthermore, the standard deviation of the CRPS over all the permutations is about an order of magnitude smaller than the improvement of the MM-EnKF compared to either the individual models or the MME.

Although for this case the assimilation order has a minimal effect, it will be important to test this sensitivity in other set-ups.

\paragraph{Partial observations}\label{ssec:partial}

We test a case where we only have partial observations of the system. In particular, here we observe only the odd-numbered $x_i$.

\begin{figure*}%[hb]
	\centering
	\includegraphics{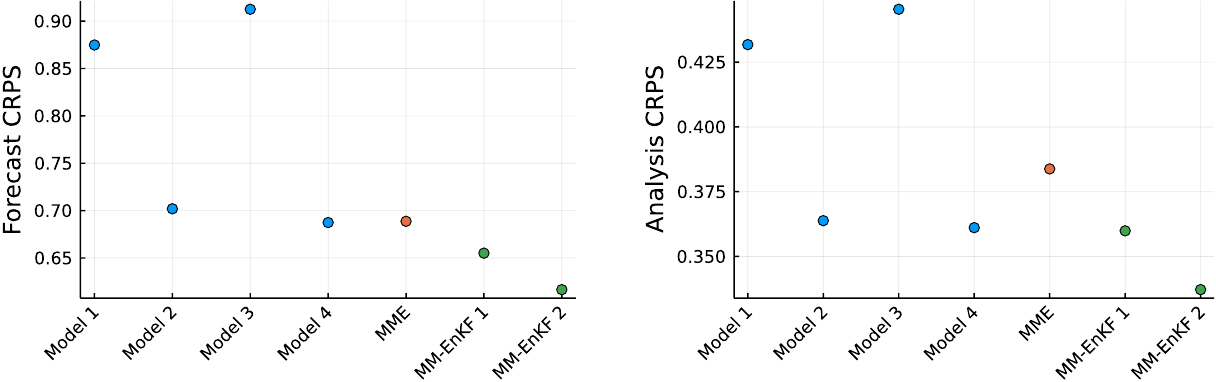}
	\caption{Overall performance of the MM-EnKF with partial observations, in terms of CRPS. The different models and multi-model combinations are indicated on the abscissa.}
	\label{fig:partial_obs}
\end{figure*}

Figure \ref{fig:partial_obs} shows the results for both forecasting and analysis. Note that the forecast step here is not different than for full observations; however, the forecasts are initialized from analyses obtained using the partial observations. Again, the MM-EnKF produces the best forecasts and analyses.

\subsubsection{Multi-model forecasts}\label{ssec:MM_Fcst}

We now test the MM-EnKF for real-time forecasting at different lead times. The experimental set-up is the same as in the previous subsection~\ref{ssec:MM_DA}, except that for each forecast cycle, we obtain the initial ensembles from a previous analysis with observations having an error of $\mathbf{R} = 0.1 \mathbf{I}$. We run 5~000 forecast cycles for each lead time, and compute the error statistics over the last 3~000 cycles.

\begin{figure}
\centering
\includegraphics[scale=0.3]{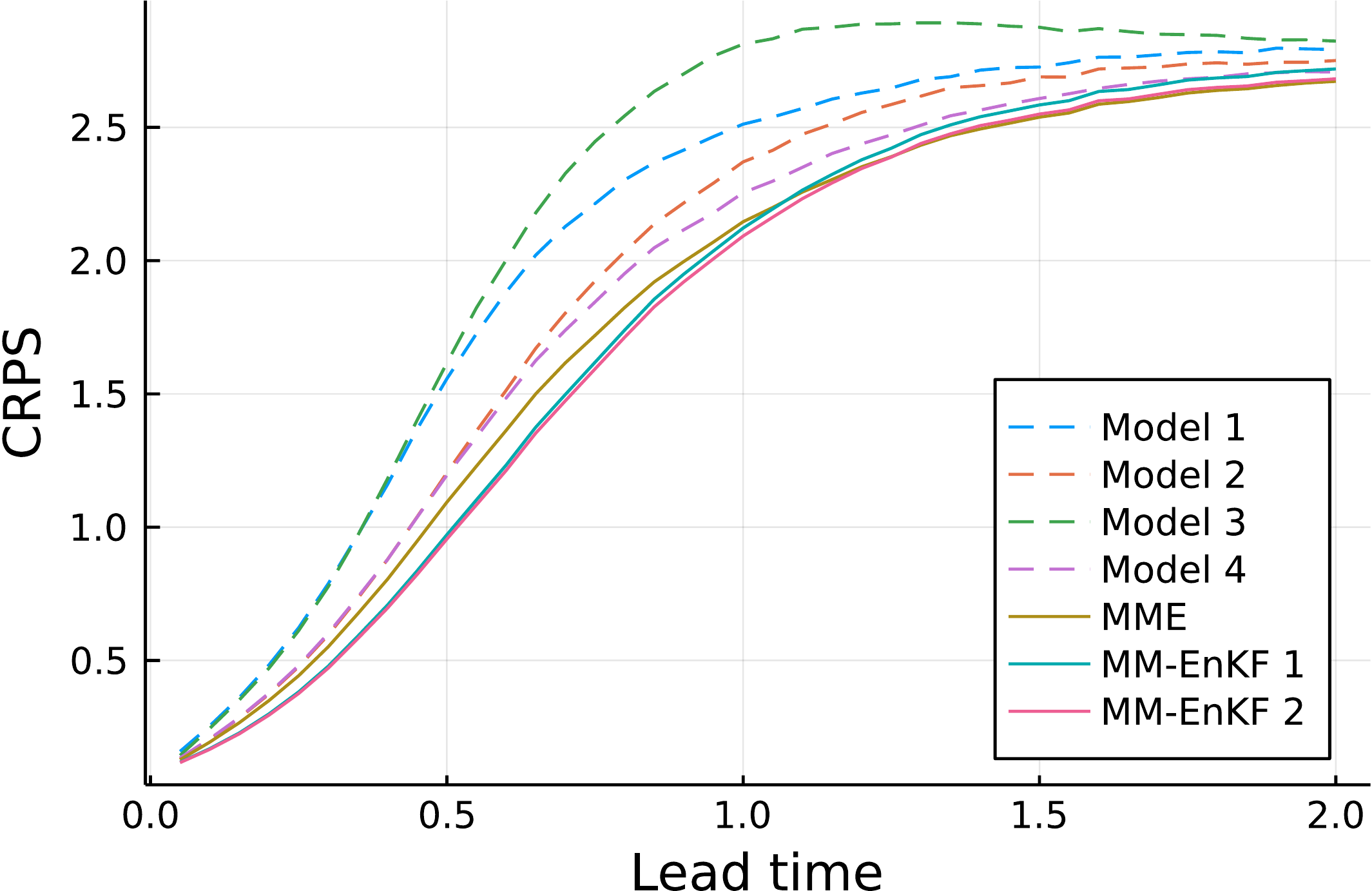}
\caption{Lead time dependence of single- and multi-model forecast performance.}
\label{fig:mm_forecasts}
\end{figure}

Figure~\ref{fig:mm_forecasts} shows that, for real-time forecasting, the MME error is similar to that of the best model, while the MM-EnKF consistently outperforms the MME and the individual models until the forecast errors start to saturate.

\begin{paragraph}{Recursive multi-step forecasts.}
We then try to apply the multi-model forecasting recursively. After an interval of 0.2, we form the multi-model forecast and use it as the initial conditions for the next interval. Figure~\ref{fig:mm_forecasts_leap} shows that this results in much greater error reductions, while Method 2 has again a slight advantage.

\begin{figure}
\centering
\includegraphics[scale=0.3]{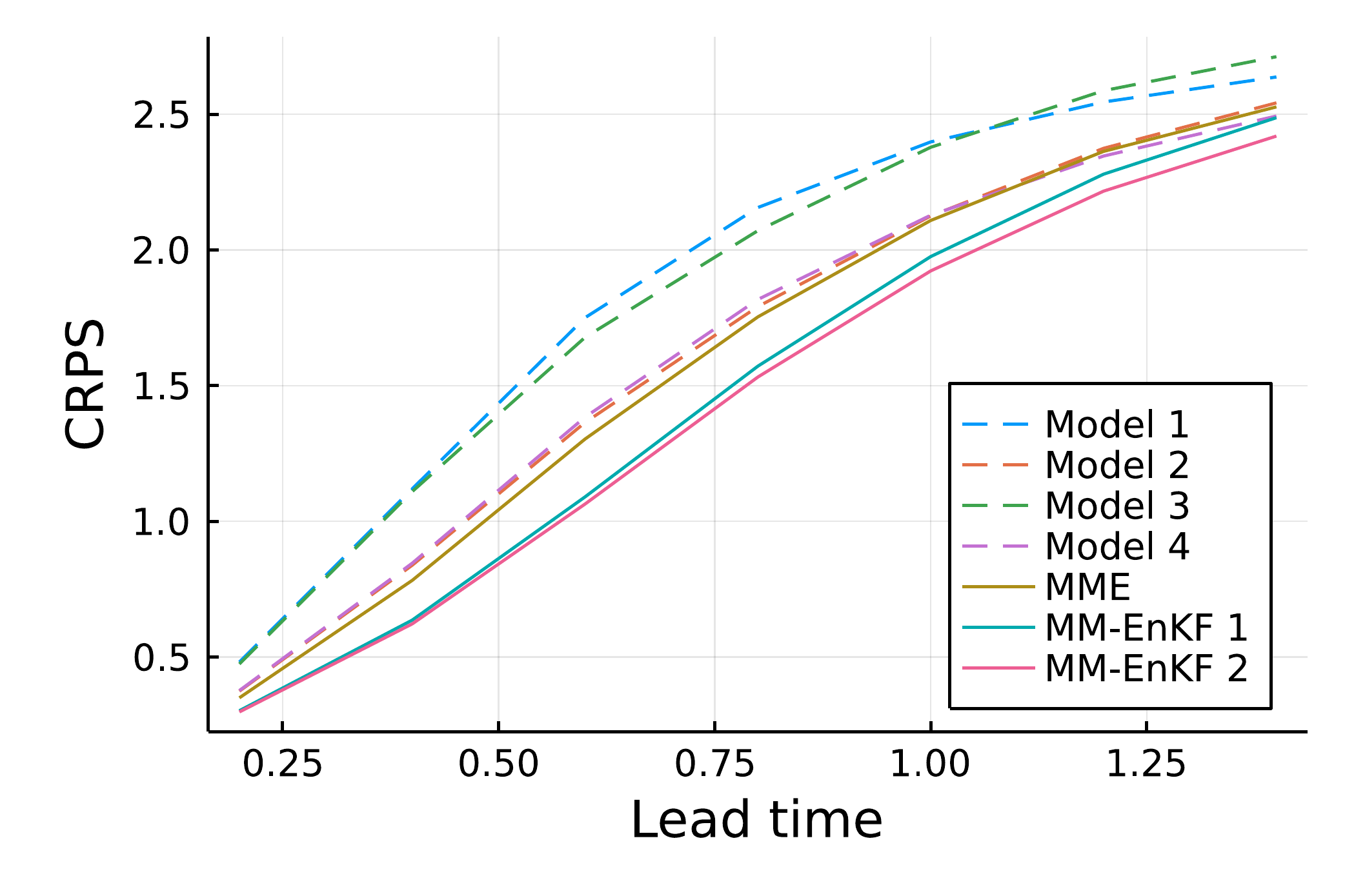}
\caption{CRPS error metric of recursive single- and multi-model forecasts by lead time.}
\label{fig:mm_forecasts_leap}
\end{figure}
\end{paragraph}

\subsubsection{Impact of flow dependence}\label{ssec:flow_dependence}

In order to estimate the impact of flow dependence in the weights, we test a 3D-Var--like version of the filter, wherein instead of using the ensemble-estimated $\mathbf{P}^\text{f}_m$ for each model, we use a static $\mathbf{B}_m$. This version is more similar to that of \cite{logutov_multi-model_2005}, which uses static forecast covariance matrices; it differs, though, from 3D-Var in that we keep the ensemble for the state update. These $\mathbf{B}_m$, instead of representing an instantaneous estimate of the forecast error covariance, represent the models' long-term statistical properties, and are often referred to as climatological error covariance matrices. We estimate these $\mathbf{B}_m$ by averaging the ensemble-estimated $\mathbf{P}^\text{f}_m$ over 100 cycles. We thus remove any flow dependence in the weights attached to the models and observations.

\begin{table*}
	\begin{tabular}{llll}
		& Analysis CRPS & Forecast CRPS & Forecast RMSE \\
		\hline
		%MME & 0.470 & 0.044 \\
		Static & $0.235 \pm 0.001$ & $0.447 \pm 0.003$ & $0.813 \pm 0.006$ \\
		Flow dependent & $\mathbf{0.202} \pm 0.001$ & $\mathbf{0.433}\pm 0.003$ & $\mathbf{0.803}\pm 0.007$ \\
	\end{tabular}
	\caption{The errors obtained for the static and flow-dependent versions of the filter. The $\pm$ indicates the standard error in the time mean. Bold indicates the lowest error in a column.}
	\label{table:flow-dependence}
\end{table*}

Comparing this non--flow-dependent version of the MM-EnKF to the flow-dependent one for both DA and forecasting in Table~\ref{table:flow-dependence}, we find that the flow-dependent MM-EnKF outperforms the non--flow-dependent version. Although the EnKF is generally known to outperform DA methods which lack flow dependence, such as 3D-Var, it is notable that the flow dependence also impacts the forecast skill. The flow dependence helps account for the uncertainty in the multi-model forecast, which is reflected in the improved CRPS. However, the flow dependence also improves the mean of the forecast ensemble, as reflected in the improved CRPS \emph{and} RMSE, the latter depending only on the ensemble mean.

\subsection{Experiments with models of different fidelities}\label{ssec:diff_ensemble_sizes}
Suppose one has two models of different accuracy and computational cost: one is more computationally expensive and more accurate, the other less expensive and less accurate. Then, can a larger ensemble of the cheaper model improve DA or forecasts of the more expensive one? Such scenarios are often encountered in operational prediction where, due to constraints on computational resources, only a small ensemble at a higher resolution can be afforded, but this can be supplemented by large low-resolution ensembles \cite{gascon_statistical_2019}. We test this scenario by applying Method 2 with models having different ensemble sizes, and additive model errors of different magnitudes.
	
We generate a $40\times 40$ banded matrix $\mathbf{B}$ with bandwidth 20; the entries within the nonzero band are drawn from a uniform distribution $\mathcal{U}(0, 1)$. We then prescribe the model error for HF (for ``high fidelity'') to have covariance $\mathbf{Q}_1 = (1/10)(\mathbf{B} - 0.4\mathbf{J}_{40})(\mathbf{B} - 0.4\mathbf{J}_{40})^T$, and LF (for ``low fidelity'') to have covariance $\mathbf{Q}_2 = (\mathbf{B} - 0.4\mathbf{J}_{40})(\mathbf{B} - 0.4\mathbf{J}_{40})^T$, where $\mathbf{J}_{40}$ is the $40\times 40$ matrix of ones.
	
We use a 5-member ensemble for HF and a 40-member ensemble for LF. In this case, the single-model forecasts do not have 45 ensemble members, rather 5 and 40, since a 45-member ensemble of HF would clearly outperform any MME which adds LF members at the expense of HF members. Rather, the question is whether the forecast skill of a small HF ensemble can be improved by adding LF members.
	
\begin{figure}
	\centering
	\includegraphics[scale=0.3]{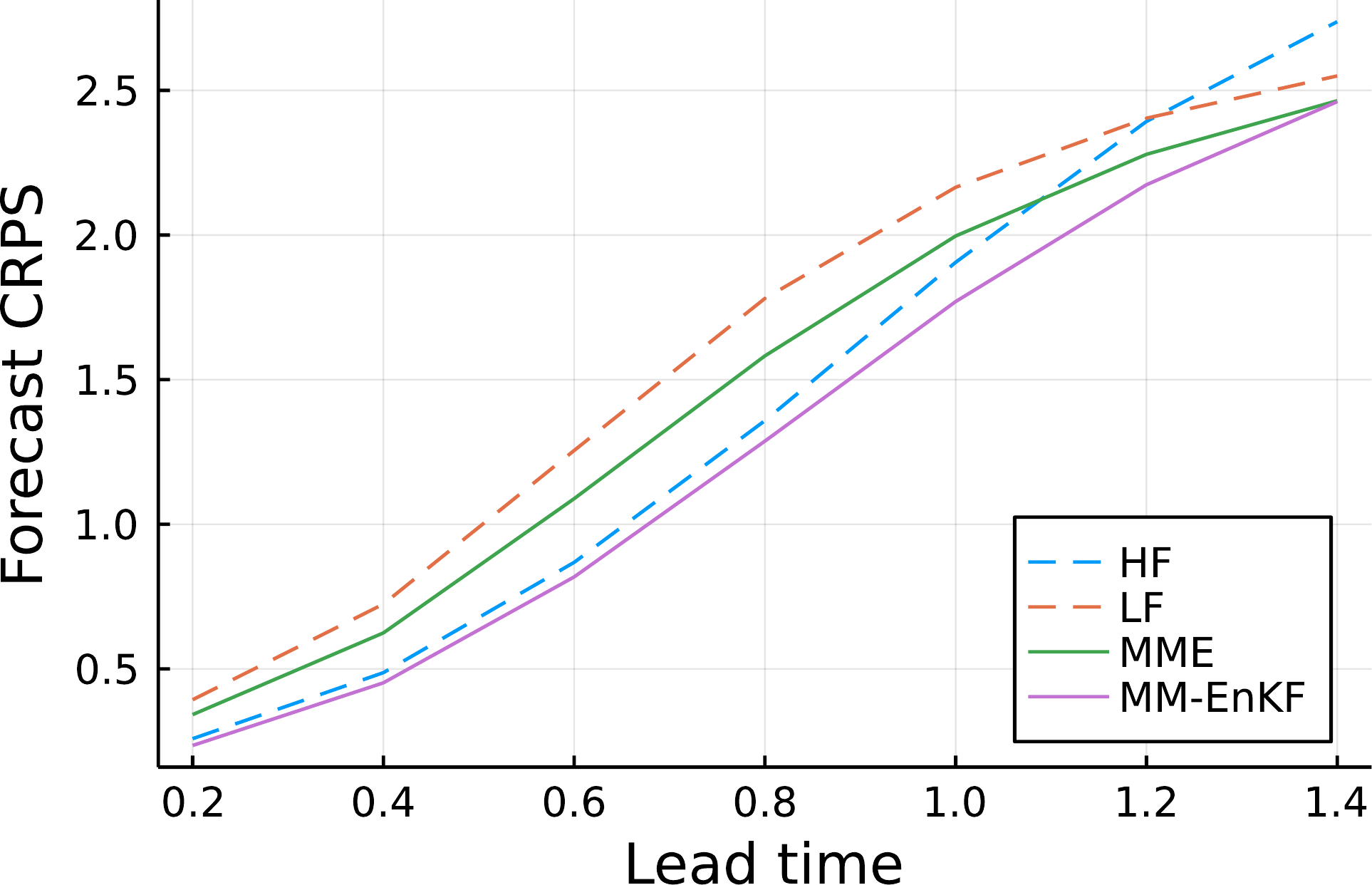}
	\caption{CRPS error metric of recursive multi-model forecasts by lead time.}
	\label{fig:crps_ml}
\end{figure}

Figure~\ref{fig:crps_ml} shows the performance of the MM-EnKF for recursive multi-step forecasts in this scenario. Here, the MME has error in between the errors of the HF and LF models, as would be expected from a simple average. On the other hand, the MM-EnKF clearly outperforms the MME and the 5-member ensemble of the more accurate HF model.
	
\subsection{Experiments with models with different resolved scales}\label{ssec:multiple_scales}
Here, we apply Method 1 with the two-scale Lorenz96 model being labeled HR (for ``high-resolution'') and the single-scale version being labeled LR (for ``low-resolution''). The two-scale Lorenz96 model includes the small-scale dynamics $\{y_j\}$ of Eqs.~\eqref{eq:Lor96} affecting the large scales $\{x_i\}$, while the single-scale version only includes the latter large scales of Eq.~\eqref{eq:LS}. This experiment thus serves as a test case for having ensembles at two different scales, one at higher resolution than the other.

The true model here is the two-scale Lorenz96 model with forcing as defined in section \ref{ssec:setup}. In this case, we prescribe an imperfect large-scale forcing of $F=8.5$ for $1 \leq i \leq 10$ and $F=9.5$ for $11 \leq i \leq 20$ for the higher-resolution model, while the lower-resolution model's forcing is perfect but model error is still present due to the unresolved scales.

Table~\ref{table:multi-scale} shows the results in terms of analysis RMSE in the large-scale variables $\{x_i\}$ and small-scale variables $\{y_{j,i}\}$. The results demonstrate that higher resolution is, at least in the present setting, more valuable than accurate forcing for the DA performance, and that the MM-EnKF provides further improvement over the better one of the two models.

\begin{table}
\begin{tabular}{lll}
& Analysis RMSE in $x_i$ & Analysis RMSE in $y_{i,j}$ \\
\hline
HR & $0.511\pm 0.007$ & $0.072\pm 0.002$ \\
LR & $0.540\pm 0.004$ & --- \\
MM-EnKF & $\mathbf{0.447}\pm 0.005$ & $\mathbf{0.066}\pm 0.001$ \\
\end{tabular}
\caption{The analysis RMSE over the large-scale variables $\{x_i\}$ and small-scale variables $\{y_{i,j}\}$ in DA experiments with the two-scale Lorenz96 model. Here the analysis window is 0.05, with 2~000 cycles and errors averaged over the last 500.}
\label{table:multi-scale}
\end{table}

\begin{figure}
\centering
\includegraphics[scale=0.3]{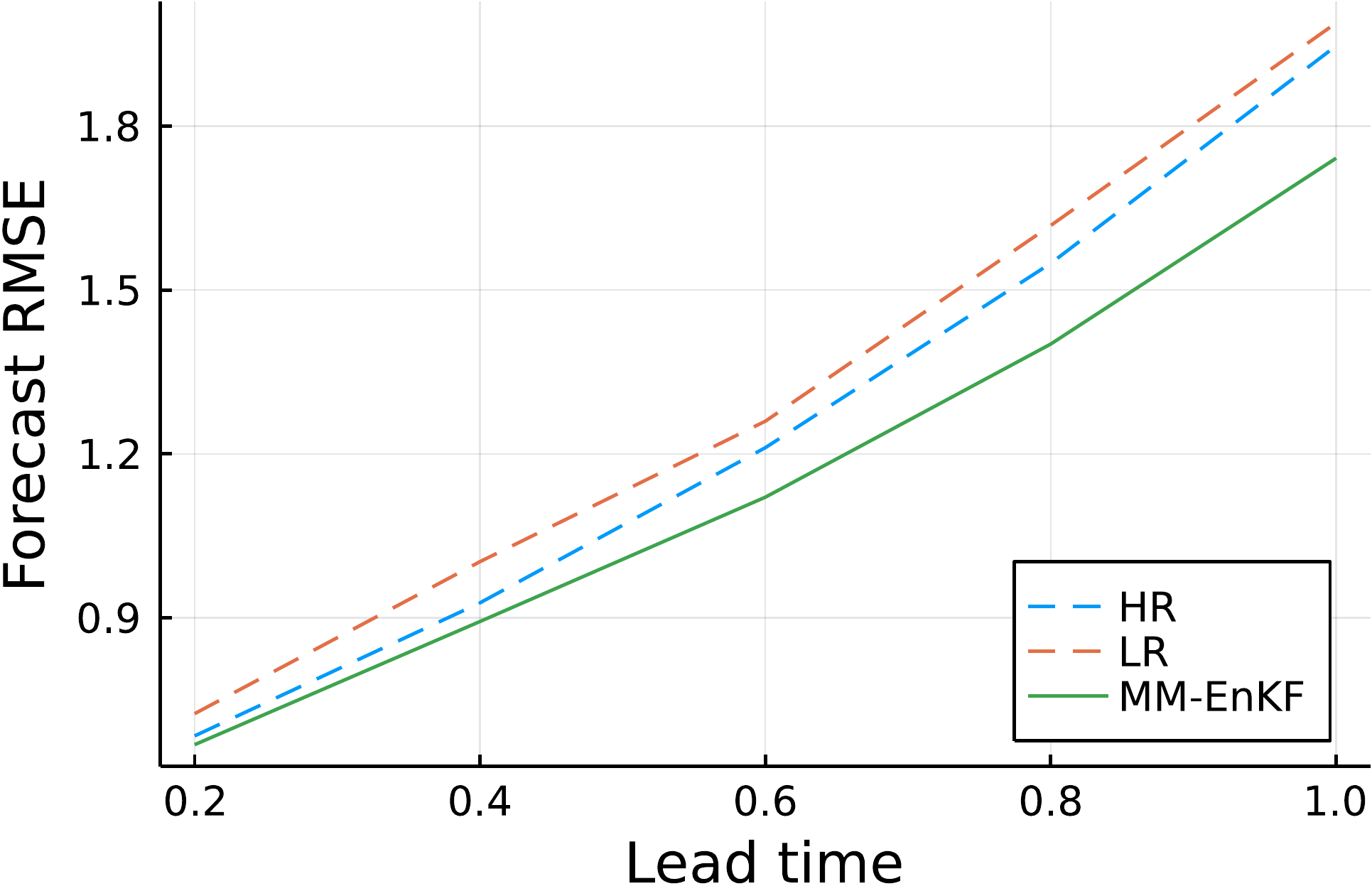}
\caption{Forecast RMSE in the large-scale variables $\{x_i\}$ by lead time.}
\label{fig:twoscale}
\end{figure}

We then test forecasting with the same two models. In these experiments, we obtain the ensembles at the beginning of each forecast cycle from a previous analysis with observations having an error of 10\% of the climatological variance. For the MM-EnKF, we forecast recursively, combining the forecasts every 0.2 time units. We run 500 cycles and show the results for the last 200 cycles in Fig.~\ref{fig:twoscale}. The MM-EnKF again outperforms both individual models by a substantial margin.

\subsection{Implementation}

We implemented the method in the Julia language, with the open-source code available (see the Open Research section). The code is modular, making it easy to add different DA methods and models. The CRPS error metric was computed with the properscoring library \cite{the_climate_corporation_properscoring_2015}. We used the parasweep library for Python of \cite{bach_parasweep_2021} to facilitate the running in parallel of multiple experiments at different lead times and with different parameter values.

\section{Concluding remarks}\label{sec:conclusions}

\subsection{Summary and conclusions}

In this paper, we proposed and implemented a multi-model ensemble Kalman filter (MM-EnKF), based on the framework of \cite{narayan_sequential_2012}. We addressed several gaps in previous work on MM-DA, including the formulation of an appropriate EnKF algorithm for high-dimensional systems and incorporation of model error estimation. Using numerical experiments with several versions of a chaotic model \cite{lorenz_predictability:_1996}, we showed that the MM-EnKF is a robust and versatile method for making use of multiple imperfect models of a system in data assimilation (DA), as well as forecasting.

\subsection{Future work}

In future work, the MM-EnKF could be applied in more high-dimensional and complex models, including operational numerical weather prediction models. Section~\ref{ssec:computation} discusses the computational considerations for such high-dimensional systems. Because the MM-EnKF can be implemented by iteratively using an EnKF, it can be adopted in contexts where an EnKF-based assimilation system already exists. It is, moreover, non-obtrusive, meaning that it does not require changes to the model equations.

In our numerical experiments, we did not perform any bias correction. However, in climate contexts, where model biases---as opposed to model error that can be approximated as unbiased noise with covariance $\mathbf{Q}_m$---become increasingly important, one will need to address the model bias issue; see section \ref{sec:model_error}. Furthermore, the interpretation of the $\mathbf{Q}_m$ becomes unclear when the ensembles mix on the models' attractors; see section \ref{ssec:forecasting}. Future work should address whether, and if so, how, the MM-EnKF can be extended to such climate problems.

Of course, the application of the MM-EnKF would also require the availability of multiple model forecasts. Examples of such operational multi-model systems include the North American Multimodel Ensemble \cite[NMME:][]{kirtman_north_2014} and the North American Ensemble Forecast System \cite[NAEFS:][]{candille_multiensemble_2009}. To our knowledge, there is not yet any operational DA systems that use multiple models. Secondarily, one would require the construction of the $\mathcal{G}_m$ operators mapping to a common space. Such operations are already used in multi-model forecast contexts, when the distinct forecasts have to be regridded to a common grid before averaging.

We discuss further avenues for future work below.

\subsubsection{Correlated forecast errors}\label{par:corr_model_err}

The formulation of MM-DA assumes that the forecast errors are uncorrelated from each other \cite{logutov_multi-model_2005}. This may not be a good assumption for climate models, especially when distinct models have the same historical provenance \cite{knutti_challenges_2010,abramowitz_esd_2019,christiansen_understanding_2020}. Future work could formulate a multi-model Kalman filter which accounts for correlated forecast errors, following Kalman filters that include correlations between observation and model errors \cite[Section 7.1 in][]{simon_optimal_2006,berry_correlation_2018}. In fact, the derivation in \ref{sec:blue_proof} is easily modified for correlated forecast error. Although for a different problem, combining correlated state estimates arises in multi-sensor fusion \cite{kim_development_1994,sun_multi-sensor_2004}, and similar ways of incorporating cross-correlation information could be tested with the MM-EnKF.

\subsubsection{Hybrid forecasting and DA}

Hybrid methods combining statistical or machine learning (ML) forecasts with a dynamical model of a system are a promising approach for improving on pure dynamical forecasts. \cite{pathak_hybrid_2018} demonstrated the advantage of hybrid models in forecasting high-dimensional chaotic systems, showing that a hybrid that combines an ML forecast with a forecast from an imperfect dynamical model can be skillful for longer than either one individually. In \cite{bach_ensemble_2021}, the authors also demonstrated the advantage of combining a data-driven model with a dynamical model in leveraging the predictability of a system's oscillatory modes. In \cite{chattopadhyay_deep_2023-1}, the authors demonstrated that augmenting an atmospheric model ensemble with a large ensemble of deep learning--based forecasts can significantly improve estimation of the forecast covariance in an EnKF.

The MM-EnKF could be tested for hybrid DA and forecasting. As demonstrated in sections \ref{ssec:diff_ensemble_sizes} and \ref{ssec:multiple_scales}, MM-EnKF is able to successfully combine models of different accuracy and resolution. This feature could be used for combining physical and data-driven forecasts: namely, one of the model ensembles could be generated by a physical model and the other one by a data-driven model.

\subsubsection{Multi-fidelity and multi-resolution forecasting and DA}

Given a limited computational budget, it appears to be advantageous to supplement a small ensemble of expensive, high-fidelity model runs with a large ensemble of cheaper, lower-fidelity runs. Future work could further explore the use of the MM-EnKF for combining ensembles at multiple fidelities or multiple resolutions. The MM-EnKF could also be tested for combining a global atmospheric forecast with multiple higher-resolution limited-area models, as in \cite{kretschmer_composite_2015}.

In addition to models that can be numerically simulated at lower resolution, another class of low-fidelity models consists of reduced-order models (ROMs) that approximate a high-fidelity model by dynamics in a lower-dimensional space. With projection-based model order reduction methods, in particular, one can project from the higher-dimensional space to the reduced space \cite{amsallem_stabilization_2012,antoulas_approximation_2005}. This idea allows one to combine high-fidelity models and ROMs within the MM-EnKF framework by defining the $\mathbf{G}_m$ operators to map from the full to the reduced space.

\subsubsection{Multi-model smoothing}

While filtering is the problem of optimally estimating the state of a system given all observations prior to the analysis time, smoothing takes into account also observations of the system obtained after the analysis time. Various forms of ensemble Kalman smoothers have been developed \cite{evensen_analysis_2018}, and future work could adapt the multi-model ensemble Kalman filter to solve the smoothing problem. For climate applications, this could enable the development of multi-model reanalyses.

\section*{Open research}
Version 2022-12 of the Julia code implementing the MM-EnKF used in this manuscript is preserved at \cite{bach_eviatarbachmm-enkf_2022}, available via the MIT License and developed openly at \url{https://github.com/eviatarbach/mmda}.

No data was used in this study. Scripts for  numerical experiments are available in the MM-EnKF repository.

\acknowledgments
We thank Marc Bocquet for several helpful suggestions, V. Balaji for discussions on correlated model error, Safa Mote for helpful discussions regarding hybrid methods, Tapio Schneider for discussions on smoothing applications, and four anonymous referees for additional suggestions. E.B. was funded by the Make Our Planet Great Again (MOPGA) postdoctoral program of the French Ministry for Europe and Foreign Affairs (no. MOPGA-977406H). The present work is TiPES contribution \#142; the TiPES (Tipping Points in the Earth System) project has received funding from the European Union's Horizon 2020 research and innovation program under Grant Agreement No. 820970. M.G. acknowledges support by the EIT Climate-KIC; EIT Climate-KIC is supported by the European Institute of Innovation \& Technology (EIT), a body of the European Union.

\appendix

\section{Optimality of the multi-model Kalman filter as the linear minimum variance estimator}\label{sec:blue_proof}
We state this result in the form of a theorem and provide its proof herewith.
\begin{theorem}
Let $\{\mathbf{\hat{x}}_\ell\}_{\ell=1}^L$ be unbiased state estimates of the $n$-dimensional vector $\mathbf{x}$ under the linear transformation $\mathbf{G}_\ell$, such that
\begin{linenomath}
\begin{equation}
\mathbf{\hat{x}}_\ell = \mathbf{G}_\ell \mathbf{x} + \mathbf{e}_\ell,
\end{equation}
\end{linenomath}
where $\mathbb{E}[\mathbf{e}_\ell] = \mathbf{0}$, $\mathbb{E}[\mathbf{e}_\ell\mathbf{e}_\ell^T] = \mathbf{P}_\ell$, and $\mathbb{E}[\mathbf{e}_\ell\mathbf{e}_{\ell'}^T] = \mathbf{0}$ when $\ell\neq \ell'$.

Then, the minimum variance linear unbiased estimator of $\mathbf{x}$ is given by
\begin{linenomath}
\begin{equation}
\mathbf{\hat{x}} = \sum_{\ell=1}^{L} \mathbf{A}_\ell \mathbf{\hat{x}}_\ell,
\end{equation}
\end{linenomath}
where
\begin{linenomath}
\begin{equation}
\mathbf{A}_\ell = \left(\sum_{\ell'=1}^L \mathbf{G}_{\ell'}^T \mathbf{P}_{\ell'}^{-1} \mathbf{G}_{\ell'}\right)^{-1} \mathbf{G}_\ell^T \mathbf{P}_\ell^{-1}.
\end{equation}
\end{linenomath}
\end{theorem}
In multi-sensor information fusion, a problem of the same form appears, except that all the $\mathbf{G}_\ell = \mathbf{I}$. For that case, minimum-variance optimality has been proven in \cite[Corollary 1 of][]{sun_multi-sensor_2004}. We largely follow the latter proof, but allow for general $\mathbf{G}_\ell$.
\begin{proof}
Begin by defining an estimator $\mathbf{\hat{x}}$ of $\mathbf{x}$ as a linear combination of the $\mathbf{\hat{x}}_\ell$:
\begin{linenomath}
\begin{equation}
\mathbf{\hat{x}} = \sum_{\ell=1}^{L} \mathbf{A}_\ell \mathbf{\hat{x}}_\ell.
\end{equation}
\end{linenomath}

Taking the expectation of $\mathbf{\hat{x}}$ and using the linearity of the expectation operator,
\begin{linenomath}
\begin{equation}
\mathbb{E}[\mathbf{\hat{x}}] = \sum_{\ell=1}^{L} \mathbf{A}_\ell \mathbb{E}[\mathbf{\hat{x}}_\ell] = \sum_{\ell=1}^{L} \mathbf{A}_\ell \mathbf{G}_\ell \mathbb{E}[\mathbf{x}].
\end{equation}
\end{linenomath}
Then, in order for $\mathbf{\hat{x}}$ to be unbiased---namely $\mathbb{E}[\mathbf{\hat{x}}] = \mathbb{E}[\mathbf{x}]$---we must have
\begin{linenomath}
\begin{equation}
\sum_{\ell=1}^{L} \mathbf{A}_\ell \mathbf{G}_\ell = \mathbf{I}.
\label{eq:weights_sum}
\end{equation}
\end{linenomath}

The error $\mathbf{\widetilde{x}}$ in $\mathbf{\hat{x}}$ can be expressed as
\begin{linenomath}
\begin{equation}
\mathbf{\widetilde{x}} = \mathbf{x} - \mathbf{\hat{x}} = \sum_{\ell=1}^{L} \mathbf{A}_\ell (\mathbf{G}_\ell \mathbf{x} - \mathbf{\hat{x}}_\ell),
\end{equation}
\end{linenomath}
with covariance matrix
\begin{linenomath}
\begin{equation}
\mathbf{P} = \mathbb{E}[\mathbf{\widetilde{x}} \mathbf{\widetilde{x}}^T] = \sum_{\ell=1}^{L} \mathbf{A}_\ell \mathbf{P}_\ell \mathbf{A}_\ell^T.
\end{equation}
\end{linenomath}

In order to obtain the minimum variance estimator, we wish to minimize $J \equiv \operatorname{tr}(\mathbf{P})$. By linearity of the trace,
\begin{linenomath}
\begin{equation}
J = \sum_{\ell=1}^{L} \operatorname{tr}(\mathbf{A}_\ell \mathbf{P}_\ell \mathbf{A}_\ell^T).
\end{equation}
\end{linenomath}
We minimize $J$ using the method of Lagrange multipliers \cite[see, e.g.,][]{boyd_convex_2004}. The Lagrangian $\mathcal{L}$ is defined as follows:
\begin{linenomath}
\begin{equation}
\mathcal{L} = J + \sum_{j=1}^{n} \left[\bm{\lambda}_j^T\left(\sum_{\ell=1}^{L} \mathbf{A}_\ell \mathbf{G}_\ell - \mathbf{I}\right)\mathbf{e}_j\right],
\end{equation}
\end{linenomath}
where $\bm{\lambda}_j = [\lambda_{1j}, \cdots, \lambda_{nj}]^T$ is the $j$th vector of Lagrange multipliers and $\mathbf{e}_j$ is a vector with a 1 in the $j$th coordinate and zeros elsewhere.
A necessary condition for $\mathcal{L}$ to have a stationary point is that
\begin{linenomath}
\begin{equation}
\frac{\partial \mathcal{L}}{\partial \mathbf{A}_\ell} = \mathbf{A}_\ell\mathbf{P}_\ell + \frac{1}{2} \mathbf{\Lambda} \mathbf{G}_\ell^T = \mathbf{0}
\label{eq:lagrange_derivative}
\end{equation}
\end{linenomath}
for all $\ell$.
We gather Eqs.~ \ref{eq:weights_sum} and \ref{eq:lagrange_derivative} into a block matrix equation:
\begin{linenomath}
\begin{equation}
\begin{pmatrix}
\mathbf{\Sigma} & \mathbf{\overline{G}}\\
\mathbf{\overline{G}}^T & \mathbf{0}
\end{pmatrix}\begin{pmatrix}
\mathbf{\overline{A}}\\
\frac{1}{2} \mathbf{\Lambda}^T
\end{pmatrix} = \begin{pmatrix}
\mathbf{0}\\
\mathbf{I}
\end{pmatrix},
\label{eq:block_matrix}
\end{equation}
\end{linenomath}
in which $\mathbf{\Sigma}$, $\mathbf{\overline{A}}$, and $\mathbf{\overline{G}}$ are the block matrices
\begin{linenomath}
\begin{equation}
\mathbf{\Sigma} = \begin{pmatrix}
\mathbf{P}_1 & & \\
& \ddots & \\
& & \mathbf{P}_M
\end{pmatrix}, \mathbf{\overline{A}} = \begin{pmatrix}
\mathbf{A}_1^T\\
\vdots\\
\mathbf{A}_M^T
\end{pmatrix}, \mathbf{\overline{G}} = \begin{pmatrix}
\mathbf{G}_1\\
\vdots\\
\mathbf{G}_M
\end{pmatrix}.
\end{equation}
\end{linenomath}

Using a block matrix inversion identity on Eq.~\ref{eq:block_matrix}, we obtain
\begin{linenomath}
\begin{equation}
\mathbf{\overline{A}} = \mathbf{\Sigma}^{-1} \mathbf{\overline{G}} \left(\mathbf{\overline{G}}^T \mathbf{\Sigma}^{-1} \mathbf{\overline{G}}\right)^{-1},
\end{equation}
\end{linenomath}
which implies
\begin{linenomath}
\begin{equation}
\mathbf{A}_\ell = \left(\sum_{\ell'=1}^L \mathbf{G}_{\ell'}^T \mathbf{P}_{\ell'}^{-1} \mathbf{G}_{\ell'}\right)^{-1} \mathbf{G}_\ell^T \mathbf{P}_\ell^{-1}.
\label{eq:opt_weights}
\end{equation}
\end{linenomath}
\end{proof}
Note that the $\mathbf{A}_\ell \mathbf{G}_\ell$ are positive semidefinite matrices. Thus, in the scalar case, Eq.~\eqref{eq:weights_sum} is a convex linear combination, i.e., the weights $\mathbf{A}_\ell \mathbf{G}_\ell$ are nonnegative and sum to 1. The multivariate case generalizes this property by having the weights be positive semidefinite matrices that sum to the identity matrix.

To apply this theorem to the assimilation step of the multi-model Kalman filter, take $L = M + 1$. Then, for $m=1,\ldots,M$, take $\mathbf{\hat{x}}_m = \mathbf{x}^\text{f}_m$ and $\mathbf{P}_m = \mathbf{P}^\text{f}_m$. Finally, take $\mathbf{\hat{x}}_{M+1} = \mathbf{y}$, $\mathbf{P}_{M+1} = \mathbf{R}$, and $\mathbf{G}_{M+1} = \mathbf{H}$. At this point, identifying $\mathbf{\hat{x}}$ with $\mathbf{x}^{\text{a}}$, we recover Eq.~\ref{eq:mm_xa} of subsection~\ref{ssec:direct}.

\section{Model error estimation method}\label{sec:Q_est}

Here, we suggest a method for estimating $\mathbf{Q}$ that is closely related to the one of \cite{berry_adaptive_2013} and \cite{hamilton_ensemble_2016}, but we assume that the observation noise covariance $\mathbf{R}$ is known. This assumption allows us to derive a simple estimate for $\mathbf{Q}$ that does not require either lagged innovations or the gain matrix. Nor is model linearization required in the case of an EnKF applied to a nonlinear forward model.

The method for estimating $\mathbf{Q}$ relies on the statistics of the innovations $\mathbf{d}(t_i) = \mathbf{y}(t_i) - \mathbf{H}\mathbf{x}^\text{f}(t_i)$, which equal the difference between observations and forecasts. A standard result for the Kalman filter states that 
\begin{linenomath}\begin{equation}
\mathbb{E}[\mathbf{d}(t_i)\mathbf{d}(t_i)^T] = \mathbf{H}\mathbf{P}^\text{f}(t_i)\mathbf{H}^T + \mathbf{R}\label{eq:desroziers};
\end{equation}\end{linenomath}
see, for instance, \cite{desroziers_diagnosis_2005} or \cite[Sec.~10.1 of][]{simon_optimal_2006}.

If the state is not fully observed, as is usually the case in DA problems, then $\mathbf{H}$ is not invertible. However, for idealized cases when $\mathbf{H}$ is invertible, we can obtain an estimate $\mathbf{\hat{Q}}$ of $\mathbf{Q}$ by substituting Eq.~\eqref{eq:pf} into Eq.~\eqref{eq:desroziers} and rearranging:
\begin{linenomath}\begin{equation}
\mathbf{\hat{Q}}(t_{i-1}) = \mathbf{H}^{-1}(\mathbb{E}[\mathbf{d}(t_i)\mathbf{d}(t_i)^T] - \mathbf{R} - \mathbf{H}\mathbf{P}^\text{p}(t_i)\mathbf{H}^T)\mathbf{H}^{-T}. \label{eq:Q}
\end{equation}\end{linenomath}
See section~\ref{ssec:rank_deficient} below for the general case in which $\mathbf{H}$ is not invertible.

In order to avoid abrupt changes in $\mathbf{\hat{Q}}$ over time, and to preserve positive semidefiniteness (see below), a temporal smoothing needs to be applied:
\begin{linenomath}\begin{equation}
\widetilde{\mathbf{Q}}(t_{i+1}) = \delta\mathbf{\hat{Q}}(t_i) + (1 - \delta)\widetilde{\mathbf{Q}}(t_i), \label{eq:mean}
\end{equation}\end{linenomath}
where $0<\delta<1$ is a tunable parameter \cite{berry_adaptive_2013,tandeo_review_2020}, and $\widetilde{\mathbf{Q}}$ is the smoothed estimate. Then, $\mathbf{P}^\text{f}(t_{i+1})$ is estimated by adding $\widetilde{\mathbf{Q}}(t_i)$ to the $\mathbf{P}^\text{p}$ estimated by the filter. In what follows, we drop the time indices for simplicity.

Covariance matrices must be positive semidefinite: in other words, their eigenvalues are real and nonnegative, i.e., $\lambda_\text{min} \geq 0$. Due to the observation noise entering the $\mathbb{E}[\mathbf{d}\mathbf{d}^T]$ term in Eq.~\eqref{eq:Q}, the estimate $\widetilde{\mathbf{Q}}$ can often lack this property. To avoid this problem, a small enough $\delta$ must be chosen, and the ``initial guess'' $\widetilde{\mathbf{Q}}(t_0)$ should be positive semidefinite. When forecasting at multiple lead times, we initialize at lead $k\tau$ by $\widetilde{\mathbf{Q}}(t_0) k^2$, inspired by the quadratic growth of model error described in \cite{carrassi_model_2008}.

In general, the larger the observation noise relative to the model error, the smaller $\delta$ must be. However, if the estimated $\widetilde{\mathbf{Q}}$ does become indefinite at some $t_j$, definiteness can be restored. The matrix satisfying $\lambda_\text{min} \geq \epsilon$ that is nearest in the Frobenius norm $\|\cdot\|_F$ \cite{horn_matrix_2013} to the problematic one at $t = t_j$ can be computed by using the spectral decomposition and setting all $\lambda_i < \epsilon$ to $\epsilon$ \cite{cheng_modified_1998}.

\subsection{Ensemble filters}

In the case of an ensemble Kalman filter, we estimate $\mathbb{E}[\mathbf{d}\mathbf{d}^T]\simeq (\mathbf{y} - \mathbf{H}\bar{\mathbf{x}}^\text{f})(\mathbf{y} - \mathbf{H}\bar{\mathbf{x}}^\text{f})^T$, where $\bar{\mathbf{x}}^\text{f}$ is the mean of the forecast ensemble.

In ensemble filters, $\mathbf{P}^\text{p}$ is estimated as
\begin{linenomath}\begin{equation}
\mathbf{P}^\text{p} = \frac{1}{m - 1} \sum_{i=1}^m (\mathbf{x}_i^\text{f} - \bar{\mathbf{x}}^\text{f})(\mathbf{x}_i^\text{f} - \bar{\mathbf{x}}^\text{f})^T,
\end{equation}\end{linenomath}
where $\mathbf{x}_i^\text{f}$ is the $i$th ensemble member and $m$ is the ensemble size. We use this $\mathbf{P}^\text{p}$ directly in Eq.~\eqref{eq:Q}, thus avoiding the need for a tangent linear model, as in Eq.~\eqref{eq:cov_prop}, when $\mathcal{M}$ is nonlinear.

\subsection{Rank-deficient observations}\label{ssec:rank_deficient}

When $\mathbf{H}$ is not invertible, we can find a solution that minimizes the Frobenius norm, as in \cite{berry_adaptive_2013}. We let $\mathbf{\hat{Q}}$ in Eq.~\eqref{eq:mean} be a linear combination of fixed matrices, $\mathbf{\hat{Q}} = \sum_p q_p \mathbf{Q}_p$. This formulation can be used to specify a simplified structure, such as a diagonal matrix or a block-constant one. 

Let $\mathbf{q}$ be the vector of coefficients $\{q_p\}$. Then,
\begin{linenomath}\begin{equation}
\mathbf{q} = \argmin_{\{q_p\}} \left\|\mathbf{C} - \sum_p q_p \mathbf{H}\mathbf{Q}_p \mathbf{H}^T\right\|_F, \label{eq:frob}
\end{equation}\end{linenomath}
where
\begin{linenomath}\begin{equation}
\mathbf{C} = \mathbb{E}[\mathbf{d}\mathbf{d}^T] - \mathbf{R} - \mathbf{H}\mathbf{P}^\text{p}\mathbf{H}^T.
\end{equation}\end{linenomath}

The minimization in Eq.~\eqref{eq:frob} is carried out by finding the least-squares solution of
\begin{linenomath}\begin{equation}
\mathbf{A}\mathbf{q} \simeq \operatorname{vec}(\mathbf{C}),\label{eq:lsq}
\end{equation}\end{linenomath}
where the $p$th column of $\mathbf{A}$ is $\operatorname{vec}(\mathbf{H}\mathbf{Q}_p \mathbf{H}^T)$.

\bibliography{mmda_arxiv}% Produces the bibliography via BibTeX.

\end{document}